\documentclass[aps,pra,10pt,twocolumn,a4paper,nofootinbib,preprintnumbers]{revtex4-1} 

\usepackage{CJK}
\usepackage{amsmath,amsfonts,amssymb,amsthm}
\usepackage{eqnarray}
\usepackage{mathcomp} 
\usepackage{srcltx}
\usepackage[normalem]{ulem} 
\usepackage{graphics,graphicx}
\usepackage{dcolumn}
\usepackage{bm}
\usepackage{hyperref}
\usepackage{url}
\usepackage[mathlines]{lineno}
\usepackage{siunitx} 
\usepackage[dvipsnames]{xcolor} 
\usepackage{units} 
\usepackage{upgreek} 
\usepackage{color, colortbl}
\usepackage{multirow, tabularx, ragged2e, booktabs, makecell} 
\usepackage{fancyhdr}

\usepackage{array} 
\newcolumntype{L}[1]{>{\raggedright\let\newline\\\arraybackslash\hspace{0pt}}m{#1}}
\newcolumntype{C}[1]{>{\centering\let\newline\\\arraybackslash\hspace{0pt}}m{#1}}
\newcolumntype{R}[1]{>{\raggedleft\let\newline\\\arraybackslash\hspace{0pt}}m{#1}}

 \newcommand{\MP}[1]{{\leavevmode\color{black}{#1}}} 

 \newcommand{\ZY}[1]{{\leavevmode\color{black}{#1}}}

\def\({\left(}
\def\){\right)}
\def\[{\left[}
\def\]{\right]}
\newcommand{\beq}{\begin{equation}}
\newcommand{\eeq}{\end{equation}}
\newcommand{\bea}{\begin{eqnarray}}
\newcommand{\eea}{\end{eqnarray}}

\newcommand{\ket} [1] {|#1\rangle} 

\newcommand{\RomanNumeralCaps}[1]{\MakeUppercase{\romannumeral #1}} 
\definecolor{LightGray}{gray}{0.9}

\graphicspath{{./Figures/}}

\begin{document}

\title{600 km repeater-like quantum communications with dual-band stabilisation}

\author{
Mirko Pittaluga$^{1,2 \dagger\ast}$,
Mariella Minder$^{1,3 \dagger}$,
Marco Lucamarini$^{1,4 \star}$,
Mirko Sanzaro$^{1}$,
Robert I. Woodward$^{1}$,
Ming-Jun Li$^{5}$,
Zhiliang Yuan$^{1}$
\& Andrew J. Shields$^{1}$
}

\affiliation{\small
\bigskip
$^{1}$Toshiba Europe Limited, 208 Cambridge Science Park, Cambridge CB4 0GZ, UK \\
$^{2}$School of Electronic and Electrical Engineering, University of Leeds, Leeds LS2 9JT, UK \\
$^{3}$Department of Engineering, Cambridge University, 9JJ Thomson Avenue, Cambridge CB3 0FA, UK \\
$^{4}$Department of Physics and York Centre for Quantum Technologies, University of York, York YO10 5DD, UK \\
$^{5}$Corning Incorporated, Corning, New York, 14831, USA \\
$^{\dagger}$These authors contributed equally to this work \\
$^{\ast}$mirko.pittaluga@crl.toshiba.co.uk\\
$^{\star}$marco.lucamarini@york.ac.uk\\
}

\begin{abstract}
\noindent
Twin-field (TF) quantum key distribution (QKD) fundamentally alters the rate-distance relationship of QKD, offering the scaling of a single-node quantum repeater.
Although recent experiments have demonstrated the new opportunities for secure long-distance communications allowed by TF-QKD,
formidable challenges remain to unlock its true potential.
Previous demonstrations have required intense stabilisation signals at the same wavelength as the quantum signals, thereby unavoidably generating Rayleigh scattering noise that limits the distance and bit rate.
Here, we introduce a novel dual-band stabilisation scheme that overcomes past limitations and can be adapted to other phase-sensitive single-photon applications.
Using two different optical wavelengths multiplexed together for channel stabilisation and protocol encoding, we develop a setup that provides repeater-like key rates over record communication distances of 555~km and 605~km in the finite-size and asymptotic regimes respectively, and increases the secure key rate at long distance by two orders of magnitude to values of practical significance.
\end{abstract}

\maketitle

\section*{Introduction}
Quantum key distribution (QKD)~\cite{BB84,Ekert.1991} allows two distant users to establish a common secret string of bits by sending photons through a communication line, often an optical fibre.
The photons, however, are scattered by the propagation medium and have only a small probability of reaching the end of the line, which restrains the QKD key rate and transmission range.
A rigorous theorem~\cite{PLOB17} (see also~\cite{TGW14}) limits to $1.44 \eta$ the number of secure bits delivered by QKD over a line with small transmission probability $\eta$, a limit known as `repeaterless secret key capacity' ($\textrm{SKC}_0$) or PLOB bound~\cite{PLOB17}.
Quantum repeaters offer a theoretical solution to extend the range of QKD~\cite{BDCZ98,DLCZ01,SSdRG11,GKF+15}.
However, a full-fledged quantum repeater remains outside the reach of present technology, due to the difficulty in building and reliably operating a low-loss quantum memory.
A partial implementation of a memory-assisted repeater has been recently achieved~\cite{BRM+20} in the form of measurement-device-independent QKD~\cite{Lo.2012} (see also~\cite{BP12}).

An alternative method to extend the transmission range of QKD without using a quantum memory has been recently discovered and named `twin-field' (TF) QKD~\cite{LYDS18} due to the peculiar interference between two fields that have related, though not necessarily identical, optical phase.
The secret key rate (SKR) of TF-QKD scales proportionally to $\sqrt{\eta}$, similar to a quantum repeater with a single node, thus entailing a major increase in the SKR-vs-distance figure of QKD.
This has led to the realisation of several experiments that display formidable long-range (or high-loss) characteristics~\cite{MPR+19,WHY+19,LYZ+19,ZHC+19,FZL+20,CZL+20}.

The security of the original TF-QKD protocol was first proved in \cite{LYDS18} for a limited class of attacks and then extended to general attacks in \cite{TLWL18} and \cite{MZZ18}.
Soon after, its experimental implementation was also considerably simplified thanks to protocol variants that waived the need for phase randomisation and reconciliation for signal states \cite{LL18,CAL19,CYW+19, WYH18,JYHW19,YHJ+19}.
The `phase-matching' protocols \cite{LL18,CAL19,CYW+19} feature signal states with a constant global phase while the `sending' or `not-sending' protocol (SNS) \cite{WYH18,JYHW19,YHJ+19} encode qubits upon optical pulses with random and unknown phases.
With the help of `two-way classical communication' (TWCC)~\cite{Gottesman.2003,Chau.2002}, the SNS protocol is able to remove the quantum bit error rate (QBER) floor intrinsic to the encoding method thereby extending the communication distance~\cite{XYJ+20}.
By running the TWCC protocol over ultralow-loss (ULL) optical fibres, a distance of 509~km has been achieved~\cite{CZL+20}, which represents the current record distance for secure quantum communications over optical fibres.

\section*{Results}

\textbf{Dual-band phase stabilisation}

In order to perform TF-QKD, it is necessary to compensate the phase drift of the encoded pulses interfering in the intermediate node (Charlie) after travelling through hundreds of kilometres in fibre.
The typical phase drift for a 100~km fibre was measured to exceed 1000~rad/s~\cite{LYDS18}.
Active compensation of rapid drift requires bright reference light to be transmitted in the same fibre along with the quantum signals for phase calibration.
The longer the fibre, the brighter the reference pulses have to be, as phase calibration requires a minimum power level to be received at the detectors.
So far, all the TF-QKD experiments used the same wavelength for both quantum and reference signals, with the help of time-divisional modulation to achieve the necessary intensity contrast.
However, this approach ceases to work for ultralong fibres.
The ever-increasing intensity of the reference pulses causes a strong Rayleigh scattering that travels back and forth along the fibre and dramatically reduces the quantum signal to noise ratio.
As proven in~\cite{CZL+20}, the noise due to double Rayleigh backscattering becomes comparable with the dark counts noise of Charlie's detectors around 500 km of ultralow-loss fibre.
Moreover, the performance of a system using a single wavelength for both dim and bright signals will be inevitably limited by the finite dynamic range of the detectors.
These two aspects fundamentally limit `single-band' TF-QKD.

In this work, we adopt a novel `dual-band' phase control using two wavelengths multiplexed on a single fibre,  which as well as solving the phase stabilisation problem in TF-QKD, could have broad applicability in a range of optical applications that require space-separated phase control.
The technique allows strong intensity contrast between reference and quantum signals while the wavelength separation prevents the Rayleigh scattering from contaminating the quantum signals.
An active phase compensation of the intense reference light leads to an immediate reduction of the phase drift by more than a factor 1000, allowing the residual drift to be compensated at a much slower pace using light signals that have comparable intensity and identical wavelength as the quantum signals.
It is worth noticing that the two wavelengths are generated by independent lasers and are not phase-locked, i.e., the stabilisation mechanism works also without an exact phase relation between the two bands.
This counter-intuitive detail is fundamental to guarantee the practicality of the setup, which makes ultra-stable cavities or complex light modulation schemes unnecessary.

The resulting setup is versatile, capable of implementing all kinds of TF-QKD protocols proposed so far, including the phase-matching ones~\cite{LL18,CAL19,CYW+19}, which cannot be efficiently run without an active phase stabilisation method.
With this setup, clocked at 1~GHz, we run various protocols and achieve record SKRs and distances for secure quantum communications over optical fibres.
The SKR overcomes the absolute SKC$_0$ at several distances, thus proving the quantum repeater-like behaviour of our system.
In addition to estimating the SKR, we also extract, for the first time, actual raw bits from a TF-QKD protocol.
This is a necessary requisite for a system that aims to distribute secure cryptographic keys to remote users in a real-world scenario.

\medskip

\textbf{Setup}

\begin{figure*}[ht!]
    \centering
    \includegraphics[width=\textwidth]{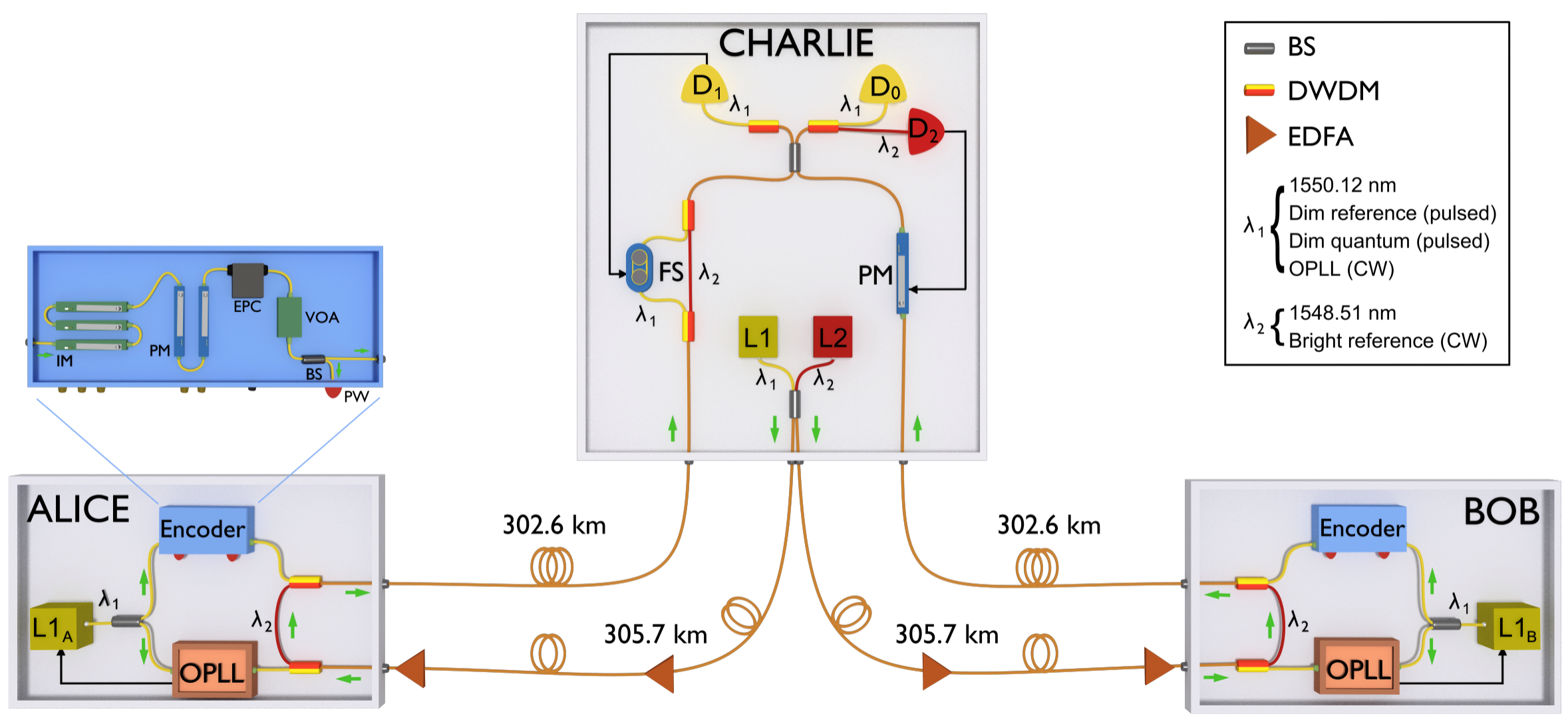}
    \caption{\textbf{Experimental setup.}
    Charlie’s L1 ($\lambda_1$) and L2 ($\lambda_2$) lasers provide continuous-wave signals for wavelength dissemination and phase tracking, respectively.
    Combined via a beam splitter (BS), they are transmitted to the symmetric users (Alice and Bob) over long servo links (305.7 km in each arm) equipped with periodic erbium doped fibre amplifiers (EDFAs).
    Each user owns an optical phase-locked loop (OPLL) to clone the $\lambda_1$ wavelength to their local lasers (L1${_A}$ and L1$_B$).
    The cloned output is encoded before being wavelength multiplexed with the disseminated $\lambda_2$ light into the quantum channel.
    Alice and Bob’s signals meet at Charlie’s second BS and interfere.
    Detectors D$_0$ and D$_1$ record the interference output for $\lambda_1$, while D$_2$ records the one for $\lambda_2$.
    The dual band phase stabilisation realised by a phase modulator (PM) and a fibre stretcher (FS) removes fast and slow phase drifts respectively.
    \textbf{Encoder boxes.}
     A set of intensity and phase modulators inside each user's Encoder allow them to run different TF-QKD protocols.
     IM: intensity modulator,
     EPC: electrical polarisation controller,
     VOA: Variable optical attenuator,
     PW: Power meter,
     DWDM: dense wavelength division multiplexer/demultiplexer.
    }
    \label{fig:setup}
\end{figure*}

The experimental setup (Fig.~\ref{fig:setup}) is composed of three modules.
The modules of Alice and Bob, who are the communicating users, transmit their quantum signals to Charlie's module via the quantum channel, made of spools of Corning SMF-28 ULL fibre.
The spools are spliced into different sets, thus enabling experiments over 5 different communication distances, ranging from 153.2 to 605.2~km.
The average loss coefficient of the fibre channel, including splices and connectors, is 0.171~dB/km. For detailed information on the fibre properties, refer to Table \RomanNumeralCaps{2} in the Supplementary Material.

The setup uses two wavelengths: $\lambda_1$ (1550.12~nm) and $\lambda_2$ (1548.51~nm), disseminated by Charlie’s L1 and L2 lasers over long servo fibre links.
Each servo link spans 305.7~km of standard single mode fibre, giving a total separation between the two communicating users exceeding 611~km.
To ensure sufficient power arriving at each user, two erbium-doped fibre amplifiers (EDFAs) are placed in each servo link to compensate for channel losses: one EDFA is placed mid-span and the other is just before the entrance to Alice/Bob.
Despite the long distance and periodic amplification, we verified the absence of detrimental nonlinear optical effects (i.e., stimulated Brillouin scattering, four-wave mixing etc.).

The users' local lasers (L1$_A$ and L1$_B$) have a free running linewidth of 50~kHz.
They are locked to the disseminated $\lambda_1$ signal through an optical phase-looked loop (OPLL), and generate light for encoding dim quantum signals.
The encoders in the users' stations operate at 1~GHz, and they carve the $\lambda_1$ input light into a train of 250~ps pulses.
The even-numbered pulses are modulated in intensity and phase, according to the requirements of the different TF-QKD protocols to be implemented.
We refer to these as `quantum signals'.
The odd-numbered pulses do not receive any further modulation and are used to track the phase drift of the quantum signals.
Hence, we refer to them as `dim reference'.
All pulses are attenuated to the single-photon level before entering the quantum channel.
A step-by-step description of the encoder modulation is given in the Methods.
The disseminated $\lambda_2$ signal is routed via dense wavelength division multiplexing (DWDM) within users’ modules for transmitting to Charlie together with the quantum signal.

Alice and Bob provide independent pre-compensation of the polarisation rotation of the signals at the two wavelengths so that all photons arrive with identical polarisation at Charlie's receiving 50/50 beam splitter (for more details on this aspect refer to Sec. \RomanNumeralCaps{3} of the Supplementary Material).
The interference output at the beam splitter are is separated by DWDM filters before detection by three superconducting nanowire single photon detectors (SNSPD's): D$_0$ and D$_1$ for $\lambda_1$  photons and D$_2$ for $\lambda_2$ photons.
Charlie's module further contains a phase modulator (PM) in one input arm and a fibre stretcher (FS) sandwiched between a pair of DWDMs in the other arm.
Full stabilisation of the quantum signal is achieved in two steps (a block diagram representation of the feedback systems is reported in Fig.~2 of the Supplementary Material), each step using a specific wavelength of the dual-band stabilisation.
First, Charlie measures the bright reference and uses a field programmable gate array (FPGA) with an integrated counter to apply a proportional-integral-differential (PID) control to the bias of his PM.
The brightness of the signal detected by D$_2$ allows this control loop to operate at 200~kHz, sufficient to stabilise the phase drift caused by the long fibre channels.
Since $\lambda_1$ and $\lambda_2$ are spectrally close and travel along the same fibre between the users and the central node, the $\lambda_2$ light can be used to stabilise the phase of the pulses at $\lambda_1$.
Assuming unidirectional phase drift, the $\lambda_2$-stabilisation will reduce the phase drift in $\lambda_2$ by approximately
$\lambda_1/\lvert\lambda_2 - \lambda_1\rvert\approx 1000$.
In the real scenario, the non-unidirectionality of the phase drift makes the actual reduction even greater, as we will show later.
The slowed drift can then be comfortably corrected through a second PID controller adjusting the bias of the FS at a rate of 10-20~Hz, without requiring an intense input signal.
The input signal for this feedback is provided by the interference outcome of the dim reference pulses at $\lambda_1$ recorded by D$_1$. More information about the feedback systems and on the sources of the residual slow drift is provided in the Methods.

\medskip

\textbf{Experimental Results}

\begin{figure}[htp]
  \centering
  \includegraphics[width=0.475\textwidth]{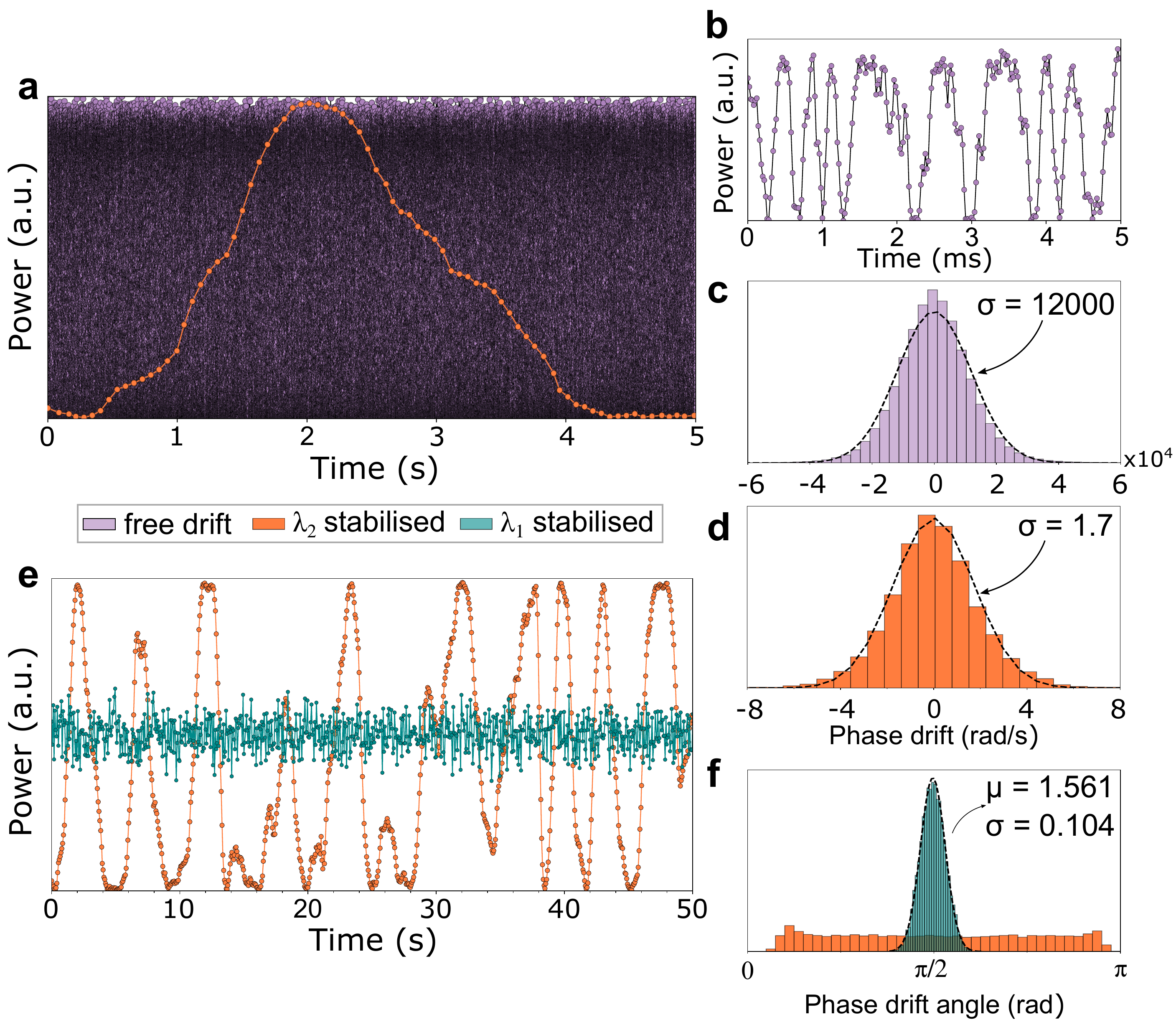}
  \caption{
  {\bf Dual-band stabilisation.}
  Data in this figure shows the interference of $\lambda_1$ \MP{light} at different stabilisation stages.
  Data was acquired over 605~km quantum and 611~km servo fibres, in a configuration identical to that in Fig.~\ref{fig:setup} except Encoder boxes were bypassed.
  Detector D$_1$ (Fig.~\ref{fig:setup}) was used to record the data.
  The colour code is: purple for free drift, orange for $\lambda_2$-stabilised data and teal for $\lambda_1$-stabilised data.
  {\bf a}, Comparison between free drifting and $\lambda_2$-stabilised data.
  Integration times were 20~$\upmu s$  and 60~ms for free drifting and $\lambda_2$-stabilised data, respectively, due to the different time scales.
  An interference visibility measurement over the free drifting ($\lambda_2$-stabilised) data yields 98.22\% (96.24\%).
  {\bf b}, Same data set as in (a) but over a ms time scale.
  {\bf c}, Histogram of the free drifting phase drift.
  The standard deviation is 11890~rad/s.
  {\bf d}, Histogram of the $\lambda_2$-stabilised phase drift.
  The standard deviation is 1.74~rad/s, i.e. about 6800 times smaller than in sub-figure (c).
  {\bf e}, Comparison between $\lambda_2$-stabilised data (orange) and data stabilised using both wavelengths, $\lambda_1$ and $\lambda_2$ (teal).
  {\bf f}, Phase offset distributions for the data shown in (e).
  $\lambda_2$-stabilised data has an almost uniform distribution over $[0,\pi]$ whereas $\lambda_1$-stabilised data has a distribution peaked around $\pi/2$.
  }
  \label{fig:feedback}
\end{figure}

Figure~\ref{fig:feedback} shows the interference outcome for $\lambda_1$, over a 605~km long quantum channel, at different stages of the stabilisation process.
The purple dots in Fig.~\ref{fig:feedback}a represent the interference when no phase stabilisation is applied.
At this distance, the free drift is so rapid (in the order of $10^4$~rad/s) that it is impossible to discern any interference fringe over a 1~s time scale.
Only on a millisecond time scale (Fig.~\ref{fig:feedback}b) we can distinguish the interference fringes.
The phase drift rate distribution associated with this measurement is shown in the purple histogram in Fig.~\ref{fig:feedback}c.
Its standard deviation allows us to quantify the phase drift in $11.89\cdot10^3$~rad/s.

After activating the stabilisation from $\lambda_2$, the phase drift rate for $\lambda_1$ reduces drastically (see orange points in Fig.~\ref{fig:feedback}a).
It is now possible to follow the evolution of constructive or destructive interference over a time scale of tens of seconds.
The effectiveness of this stabilisation is quantifiable by the reduction in the phase drift rate for the recorded data (orange histogram in Fig.~\ref{fig:feedback}d).
When feedback from the bright reference at $\lambda_2$ is enabled, the standard deviation of the drift rate decreases to $1.74$~rad/s, a value $\sim$6800 times smaller than without the bright reference stabilisation.
This reduction is considerably better than the estimated factor 1000 due to the cancellation of rapid opposite drifts.
The residual slow phase drift of $\lambda_1$ can be readily compensated by using the dim reference pulses at this wavelength, which leads to a stable interference output (teal dots in Fig.~\ref{fig:feedback}e).
Figure~\ref{fig:feedback}f shows the phase distribution between the interfering $\lambda_1$ signals locked to have $\pi/2$ difference.
The locking error is only 0.10~rad (standard deviation of the teal coloured distribution in the figure), which contributes to the QBER by approximately 2\%.

\begin{figure}[ht!]
  \centering
  \includegraphics[width=\columnwidth]{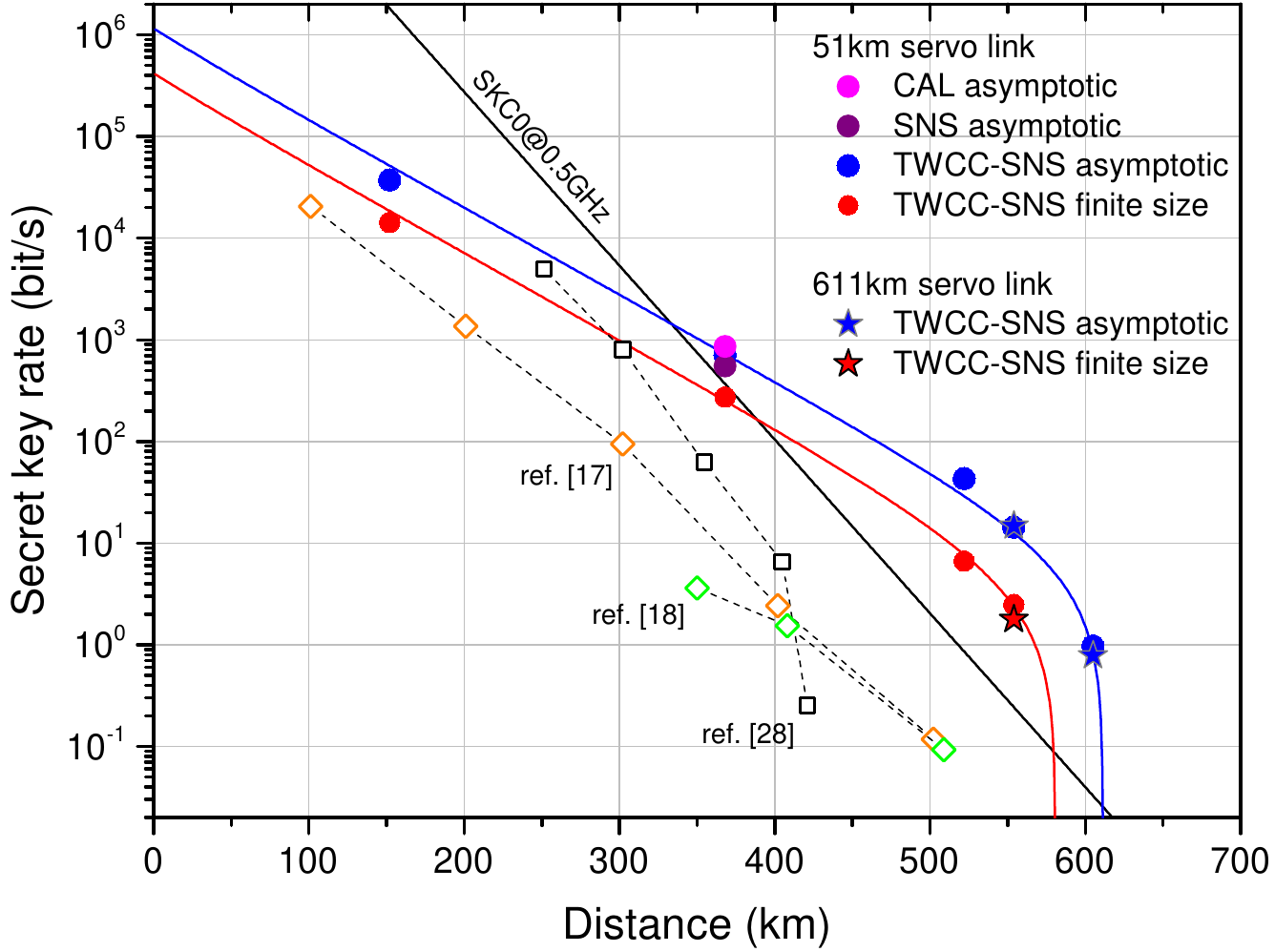}
    \caption{\textbf{Key rate simulations and results.}
    Secret key rates are plotted against the quantum channel length.
    This is constituted by ultra-low loss (ULL) fibres of 0.171~dB/km loss.
    The SKC$_0$ bound for unitary detection efficiency (black line) is plotted along the simulations for the TWCC SNS TF-QKD protocol in the asymptotic and finite size regimes (blue and red curves respectively).
    Filled markers show the experimental results we obtained for the different protocols whereas unfilled markers are the state-of-the-art results in term of SKR over distance for fibre-based TF-QKD \cite{FZL+20,CZL+20} (diamonds) and QKD \cite{BBR+18} (squares).
    }
  \label{fig:SKR}
\end{figure}

Using the described dual-band stabilisation, we performed four experiments with different TF-QKD protocols, varying the operational regimes and optimising the parameters in each case.
Firstly, the CAL~\cite{CAL19} and SNS~\cite{WYH18} protocols in the asymptotic regime, then the SNS with the TWCC method~\cite{XYJ+20}, both in the asymptotic and in the finite-size regimes~\cite{JYHW19,YHJ+19}.
In the practically relevant case of finite-size TWCC-SNS, we also extracted real bits of the raw key.
We performed these experiments in two stages. First, we developed a simplified asymmetric setup to assess the feasibility of long-distance TF-QKD with dual-band phase stabilisation, featuring a single OPLL and a 51~km servo fibre.
We then moved on to a symmetrical configuration where the frequency reference is disseminated by Charlie (Fig.~\ref{fig:setup}) via 611~km servo fibre for the final experiments over the two longest quantum channel fibre distances.
Details about the asymmetric experimental setup, the protocol parameters, together with additional information on the patterns used for encoding, are given in Sec. \RomanNumeralCaps{1} and \RomanNumeralCaps{5} of the Supplementary Material.

In Fig.~\ref{fig:SKR} we report our results in terms of SKR versus distance, together with the simulation curves and the state-of-the-art SKRs for long-distance TF-QKD \cite{FZL+20,CZL+20} and QKD \cite{BBR+18} over optical fibres.
In the same graph we also plot the absolute SKC$_0$, which assumes ideal equipment for Alice and Bob and hence is the most difficult bound to overcome.
Surpassing this limit proves the repeater-like behaviour of our setup.
The complete experimental results can be found in Sec. \RomanNumeralCaps{6} of the Supplementary Material.
The CAL and SNS protocols have been implemented on a 368.7~km-long optical fibre (62.8~dB loss) and analysed in the asymptotic scenario.
For CAL, we obtain an SKR of 852.7~bit/s, 2.39 times larger than SKC$_0$.
For SNS, the SKR is 549.2~bit/s, 1.54 times larger than SKC$_0$.

Using the TWCC SNS version of TF-QKD, we take measurements at 153.3, 368.7, 522.0, 555.2 and 605.2~km, i.e., from 26.5~dB to 104.8~dB loss, and we extract positive SKRs both in the asymptotic and in the finite-size regimes.
In Fig.~\ref{fig:SKR}, blue (red) symbols refer to the experimental results obtained in the asymptotic (finite-size) case scenario.
Stars (dots) represent results obtained through the symmetric (asymmetric) setup with a 611~km (51~km) servo fibre.
Despite periodic optical amplifications, the longer servo link introduces only a marginal reduction of the secret key rate.
At a 555~km quantum channel and a 611~km servo link, with less than 2~h of continuous measurement, we are able to extract a finite-size SKR of 1.777~bit/s, a value 7.68~times higher than the absolute SKC$_0$.
Extending the quantum channel to 605.2~km, with a loss budget of 104.8~dB, we achieve an asymptotic SKR of 0.778~bit/s, which is 24 times higher than the SKC$_0$.
This represents the first fibre-based secure quantum communication beyond the barriers of 600~km and 100~dB.

To further appreciate the progress entailed by our new technique, we compare our results with the experimental points setting the current record distances for fibre-based QKD (421~km \cite{BBR+18}) and TF-QKD (502~km \cite{FZL+20}, 509~km \cite{CZL+20}).
Distance-wise, there is an increase of tens of (more than a hundred) kilometres over TF-QKD (QKD) prior art.
The main element enabling the distance improvement over previous TF-QKD implementations is the dual-band stabilisation technique, which leads to negligible contamination of the encoded signal by the bright reference.
In previous experiments, the bright stabilisation signal was emitted at the same wavelength as the encoded signal, thus causing an intense double Rayleigh backscattering that ultimately limited the maximum communication distance.
In our case, on the other hand, even at the longest distance the noise introduced by the stabilisation signal was below the detectors' dark counts.

The dual-band stabilisation technique leads also to an even more pronounced enhancement of the SKR, with an improvement of 2 orders of magnitude at 500~km, the furthest distance achieved by prior art.
This is possible because we could keep the clock rate of the encoded signals at the high value of 500~MHz at all distances.
In previous experiments, where the stabilisation signal was time-multiplexed, the protocol clock rate had to be reduced considerably to accommodate for reference signals, and to leave some recovery time at the detectors (after these received the bright intensity reference pulses).

\begin{figure}[b]
  \centering
  \includegraphics[width=\columnwidth]{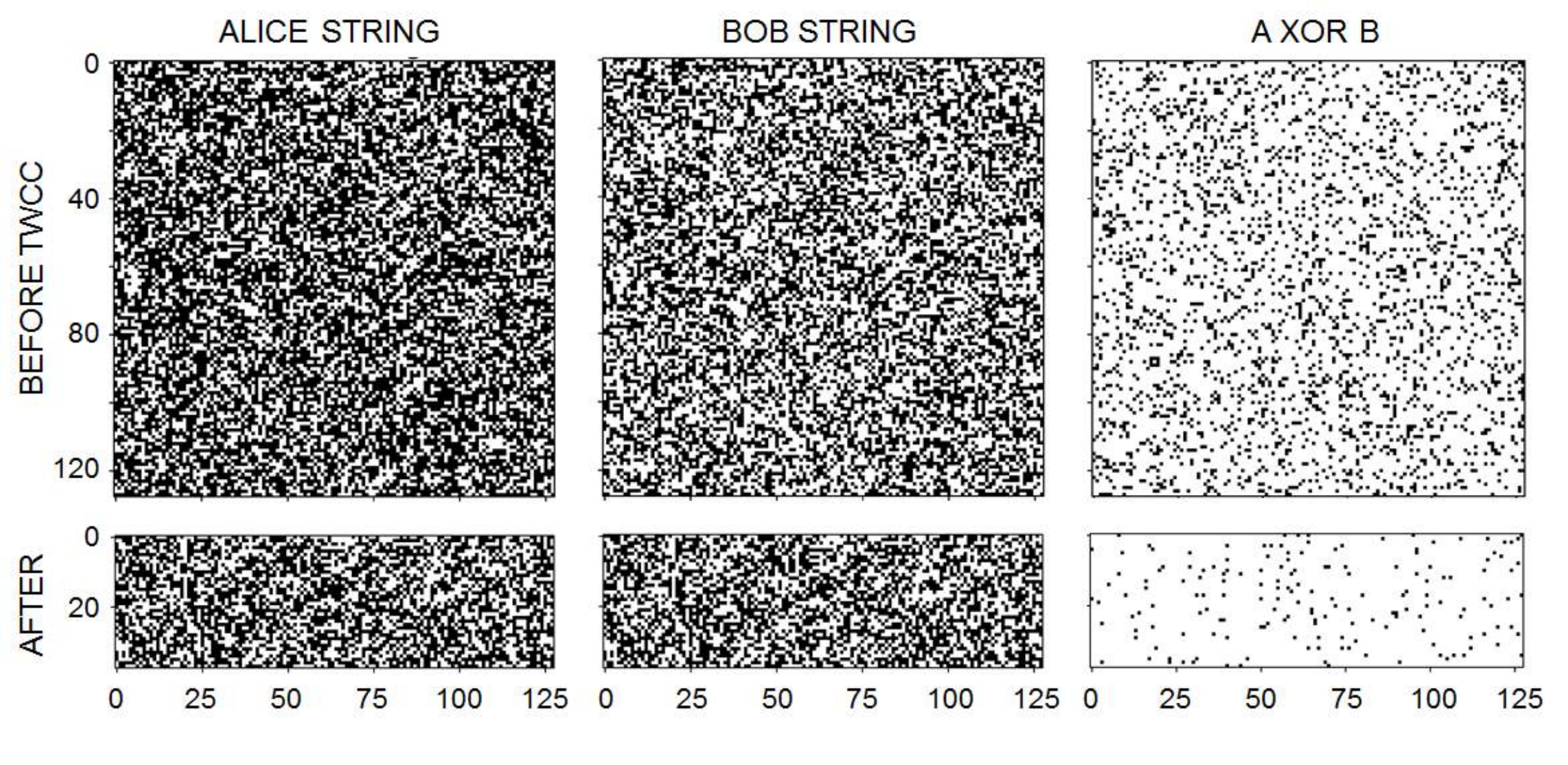}
 \caption{
    \textbf{Binary maps of the extracted bit strings.}
 	Samples of the bits extracted from the experiment performed at 522~km before (top panels) and after (bottom panels) TWCC is applied.
 	\textit{Top:}~The first two squares on the left ($128\times128$ pixels) are a sample of the users' raw strings before TWCC is applied, with white (black) pixels associated with the bit value 0 (1).
    The third square on the right is obtained by modulo-2 addition (XOR) of the first two.
    The black dots in this square represent the errors in the strings.
 	\textit{Bottom:}~Refined keys after TWCC has been applied.
    The strings shrink by 70\% into rectangles with $128\times38$ pixels.
    Reduction in key size is accompanied by a substantial reduction in the key errors, as is apparent from the rightmost rectangle.
    }
  \label{fig:2WCCraw}
\end{figure}

All the TF-QKD experiments performed so far, as well as the vast majority of long-distance QKD experiments, have only provided an in-principle estimation of the SKR without a real extraction of the bits that form a cryptographic key after suitable post-processing.
In our experiment, we extract real strings of bits from the SNS protocol and process them with the TWCC method.
The generation of raw bits is a challenging task, especially with a clock rate as high as 1~GHz, as it requires individual tagging and real-time manipulation of the signals recorded at the detectors.
Figure~\ref{fig:2WCCraw} gives a graphical representation of the TWCC method applied to a raw bit string extracted during the experiment performed at 522~km.
The bits of the strings are displayed as white or black pixels depending on their value 0 or 1, respectively.
The leftmost and central panels in the first row show Alice's and Bob's raw strings, distilled from the SNS protocol, whereas the rightmost panel reports the bitwise addition of the two strings.
The density of the dots in the first two panels reveal a slight bias (53.8\%) in the bit value which is intrinsic to the SNS protocol \cite{WYH18}.
A simulation shows that with our parameters a bias of 52.7\% has to be expected.
On the other hand, the black dots in the rightmost panel highlight the conflicting bits in the users' raw keys, which amount to a 16\% of the total.
The second row of Fig.~\ref{fig:2WCCraw} shows the effect of TWCC on the users strings.
TWCC induces a considerable reduction of the errors, from 16\% to 3.5\%, and of the bias in the strings at the expenses of the strings length, which decreases by $\sim$70\%.
However, the overall effect of TWCC is beneficial, as it increases the signal-to-noise ratio of the raw keys and so also the range of TF-QKD.

\section*{Discussion}

We have shown that dual-band phase stabilisation can dramatically reduce the phase fluctuations on optical fibre by almost four orders of magnitude.
This has allowed us to overcome the fundamental noise limitation of long distance TF-QKD and increase its secret key rate from the current millibit per second range to the bit per second range for the longest fibre length.
We notice here that 1 bit/s key generation rate is sufficient to enable fast key refresh of symmetric cryptographic protocols, such as AES, several times per day.
Our setup tolerates a maximum loss beyond 100 dB allowing quantum communication over 600~km of fibre for the first time.
We believe these techniques will have more general application in quantum communications, for example enabling DLCZ-type quantum repeaters~\cite{DLCZ01}, longer-baseline telescopes~\cite{Gottesman.2012}, quantum fingerprint~\cite{AL14,XAW+15,ZXLQ20} over longer distances or a phase-based architecture for the quantum internet~\cite{Kim08}.

\emph{Note added -}
During the completion of our work, one of the anonymous referees noted that the finite-size equations we borrowed from Refs.~\cite{JYHW19,YHJ+19} only hold if the variables are i.i.d., a result not known at the time of writing. A full analysis of this point has only recently appeared in a preprint~\cite{JHY+21} and suggests that the removal of the i.i.d. assumption only entails a slight increase in the failure probability of the protocol.

\begin{acknowledgments}
We thank Xiang-Bin Wang and Hai Xu for their useful feedback on the TWCC protocol.
The authors acknowledge funding from the European Union’s Horizon 2020 research and innovation programme under the grant agreement No 857156 ``OPENQKD'' and under the Marie Sk\l{}odowska-Curie grant agreement No 675662.
M.M. acknowledges financial support from the Engineering and Physical Sciences Research Council (EPSRC) and Toshiba Europe Ltd.
\end{acknowledgments}

\medskip
\textbf{Author contributions.}
M.P. and M.M. developed the experimental set-up, performed the measurements and analysed the data.
M.S. and R.I.W. supported the experimental work.
M.-J.L. provided the ultralow-loss fibres.
Z.L.Y., M.L. and A.J.S. guided the work.
M.L., M.P. and M.M. provided the simulations and wrote the manuscript, with contributions from all the authors.

\medskip
\textbf{Competing interests.} The authors declare no competing interests.

\medskip
\textbf{Data availability.} The data that support the plots within this paper and other findings of this study are available from the corresponding authors on reasonable request.

\medskip
\textbf{Code availability.} The codes used to process the data for this paper are available from the corresponding authors on reasonable request.

\medskip
\textbf{Correspondence and requests for materials.} should be addressed to M.P. and M.L.


\clearpage
\small 
\section*{Methods}
\textbf{Encoder boxes.}
For a detailed representation of the components inside the encoder boxes see inset diagram in Fig.~\ref{fig:setup}.
The incoming CW light arrives already aligned in polarisation with the optical axes of the subsequent modulators.
The first components in the encoders are three intensity modulators~(IMs), used to carve 250~ps long pulses at a 1~GHz rate, with three possible intensity levels ($u$, $v$, $w$).
The intensity ratios between the different intensity levels can be adjusted by the AC amplitude driving the IMs.

Two phase modulators~(PMs) are then used to encode the phase of the optical pulses.
In this system, we cascade two PMs instead of using just one to reduce their RF signal amplitudes.
Limiting each PM to a modulation range of $[-\frac{\pi}{2},\frac{\pi}{2})$, we achieve a phase modulation that covers the whole $[0,2\pi)$ range and that is linear with its driving signals amplitude.
Each PM is driven by a 8-bit DACs, and with two cascaded we are able to encode 512 different phase values over the $2\pi$ phase range.

All the modulators are driven by two synchronised 12~GSa/s waveform generators, one for each user, programmed to encode a 25040-pulse long pseudo-random pattern.
For more information on the encoded pattern refer to Sec. \RomanNumeralCaps{6} of the Supplementary Material.

The PMs are followed by an electrically driven polarisation controller~(EPC), a variable optical attenuator~(VOA), and a 99:1 beam splitter~(BS).
The EPC is used to control the polarisation of the $\lambda_1$ photons after transmission through the channel.
Each user has a continuous polarisation optimisation routine that aligns the quantum signals along the preferred optical axis at Charlie.

The VOA sets the flux of the quantum signal before injection into the quantum channel, through a flux calibration control loop that continuously adjusts the VOA so as to have a stable optical output, monitored at the strong output of the BS.

\medskip

\textbf{Feedback systems.}
The dual-band phase-stabilisation strategy employed in this experiment enabled us to stabilise the quantum channel without affecting the encoding in the wavelength reserved for the quantum signal ($\lambda_1$) or the clock rate of the protocol, which was kept at 500~MHz at all the tested distances.
Its general design is presented in Fig.~\ref{fig:setup} and its detailed block diagram is given in Fig.~2 of the Supplementary Material.

There, Fig.~2a shows the stabilisation method based on the bright reference at $\lambda_2$.
It features a closed loop cycle that locks the interference between Alice's and Bob's bright reference beams to a given intensity level.
This, in turn, locks the phase offset between these signals to a fixed value.
The bright reference interference is monitored by the SNSPD $D_{2}$.
Single photons detected by $D_{2}$ are integrated over a period of 5~$\upmu s$.
The difference between the integrated number of counts and the set value, constitutes the error signal of a PID controller implemented with an FPGA clocked at 200 kHz.
By tuning the DC offset of a phase modulator (PM) that acts on the light coming from Bob, the FPGA controls the interference between the bright references.
It is important to notice here that the phase shift applied by the PM affects both the wavelengths $\lambda_2$ and $\lambda_1$.
The feedback based on $\lambda_2$ fully stabilises the bright reference light while it only partially stabilises the quantum one.

The remaining (slow) phase drift on $\lambda_1$ is related to two factors: the fact that $\lambda_1$ and $\lambda_2$ travel separately in certain sections of the setup (necessary for the protocol encoding over $\lambda_1$ at the transmitting stations), and the fact that the fast feedback introduces a phase drift over $\lambda_1$ when the length difference between the two channels varies over time.
The former component of the slow phase drift can be seen as the phase noise picked up by an asymmetric Mach-Zender interferometer having the dimensions of those sections of the setup where the two wavelengths travel separately.
The latter component can be explained as a consequence of the finite range of the PM, and of the phase locking of the fast feedback over $\lambda_2$, rather than $\lambda_1$.

The PM in the fast feedback actively compensates the fast phase drift.
However, its finite adjustment range is incapable of compensating at entirety the phase drift caused by fibre length variation.
It must rely on multiple ($M$) resets in order to maintain the $\lambda_2$ phase difference to $\phi = 2 \pi M + \phi_t$, where $\phi_t$ is the target phase.
Due to the $\lambda_2 - \lambda_1$ wavelength difference, this compensation will introduce a residual phase drift ($\Delta \phi$) over $\lambda_1$ equal to:

\begin{equation}\label{Residual_Phase_Drift}
\Delta \phi = 2 \pi M \cdot \left( \frac{\lambda_2 - \lambda_1}{\lambda_1} \right).
\end{equation}

The residual drift introduced by the $\lambda_2$-stabilisation over $\lambda_1$ is estimated to be $\frac{\phi}{\Delta \phi} = \frac{\lambda_1}{\lambda_2-\lambda_1} \approx 1000$ times smaller than the original fibre phase drift, if assuming unidirectional fibre length drift.
In reality, the fibre length drift direction is random.
With cancellation of positive and negative $2\pi$ resets, we obtain experimentally a higher reduction factor of $\sim 6800$ (as shown in Fig.~\ref{fig:feedback}).

Supplementary Material Fig.~2b shows the stabilisation mechanism that corrects the residual phase drift on $\lambda_1$.
The error signal for it is provided by the overall interference of quantum signals and dim reference.
The quantum signals are interleaved with the dim reference pulses, which are unmodulated and have the same intensity as the brightest decoy pulse ($u$).
The presence of dim reference pulses guarantees that the averaged output of the interference is directly related to the residual phase offset in $\lambda_1$.
This is retrieved by integrating the single photons detected by SNSPD $D_1$ over 50~ms or 100~ms, depending on the distance.
The difference between this value and a set value provides the error signal for a PID controller implemented with a micro-controller operating at the frequency of 20~Hz or 10~Hz, depending on the distance.
The micro-controller corrects the phase offset by modulating a fibre stretcher acting on the quantum signal coming from Alice.
Differently from the stabilisation in $\lambda_2$, the one in $\lambda_1$ acts solely on the quantum signals and can therefore correct its residual phase drift.

Due to the different expansion/contraction rates of the channels connecting Charlie to the two users, during the protocol execution we had to compensate for the change in length of the quantum channels.
We did that by opportunely delaying the pattern encoding of one user with respect to the other, aiming at obtaining always optimal time alignment of the users' pulses at Charlie's BS.
The intervals between these alignment adjustments depended on the stability of the environmental conditions in the lab, and varied from once every 4 minutes, up to once every of 30 minutes.
From the highest adjustments frequency, we estimated an upper limit of the length difference drift between the two sides of the communication channel (in our air-conditioning temperature stabilised lab) of $\sim$3~mm/min in the longest experimental setting.

\medskip

\textbf{Protocols.}
To demonstrate the multi-protocol aspect of our system, we implemented different variants of TF-QKD, in different regimes.
We list them as CAL~\cite{CAL19}, SNS \cite{WYH18,YHJ+19,JYHW19} and TWCC-SNS \cite{XYJ+20}.
Their detailed description and security proofs can be found in the referenced papers.
See also Methods in \cite{MPR+19}.
Here we describe our encoding method and the equations used to extract the secret key rate from each protocol.

In all protocols, we consider a symmetric situation, with identical photon fluxes for the users Alice and Bob.
This is the real situation in the experiment, where fibre lengths and losses between the users and Charlie are nearly identical (see e.g. Table \RomanNumeralCaps{2} in Supplementary Material).
Therefore we only describe the relevant steps for the user Alice; Bob will execute similar operations in his own location.
During the preparation stage, Alice generates weak coherent states of the form $\ket{\sqrt{\mu} e^{i \theta}}$.
She randomly selects a basis $X$ or $Z$ with probabilities $P_X$ or $P_Z$ ($P_X+P_Z=1$).
If she chooses $X$ (test basis), she randomly selects a flux value $\mu = \{u, v, w\}$ with conditional probability $P_{\mu|X}=\{P_{u|X},P_{v|X},P_{w|X}\}$, $P_{u|X}+P_{v|X}+P_{w|X}=1$, and a random global phase value $\phi\in[0,2\pi)$.
She then prepares and send the phase-randomised weak coherent state $\ket{\sqrt{\mu}e^{i \phi}}$.
If she chooses $Z$ (code basis), she randomly selects a bit value $\alpha = \{0, 1\}$ and sets the photon flux to $\mu = \{s,n\}$ with conditional probability $P_{\mu|Z}=\{P_{s|Z},P_{n|Z}\}$, $P_{s|Z}+P_{n|Z}=1$.
In CAL, bits are encoded as coherent states $\ket{\sqrt{s}e^{i \alpha \pi}}$.
In SNS, bits are encoded on the photon flux, with $s$ ($n$) representing a bit value 1 (0) for Alice and a bit value 0 (1) for Bob.
With our encoder, the photon fluxes $w$ and $n$ are both very small, in the order of $10^{-4}$.
Therefore sending out a photon flux $n$, or $w$, is equivalent by all practical means to not sending out any flux at all.
We denote the probability of `not sending' conditional on choosing the $Z$ basis as $P_{n|Z}$ and the probability of sending a photon flux $s$ conditional on the $Z$ basis as $P_{s|Z}$ or simply $\epsilon$.
The detailed values of the parameters used in the experiment depend on the protocol (CAL, SNS, TWCC) and on the regime (asymptotic or finite-size) adopted.
They are listed in Tables \RomanNumeralCaps{3} and \RomanNumeralCaps{4} in the Supplementary Material.

After the preparation stage, Alice and Bob send their pulses to central node, Charlie.
Charlie should interfere the received pulses on a beam splitter and measure the result, announcing publicly which detector click.
If Charlie is malicious and adopts a different detection and announcement strategy the security of TF-QKD remains unaffected.
After a total of $N_{0}$ signals have been sent, the quantum transmission is over and Charlie publicly announces his measurements.
When Charlie's announcement is complete, the users announce their bases.
For the $X$ basis, they also disclose their intensities $\mu$ and, limitedly to the SNS protocol, they announce the values of their global phases $\phi$.
%
Alice and Bob post-select the events for which they used matching bases and intensities.
For SNS, they also select the events with global phase values not mismatched by more than $\Delta$ modulo $\pi$ .
The users extract the bits from the $Z$ basis events and use the $X$ basis events to perform the security analysis.
In TWCC, the bits in the string distilled from the $Z$ basis are randomly paired and bit-wise XOR-ed.
More specifically, Bob randomly pairs the bits up and announces the positions and parities of each pair.
Alice uses this information to repeat this step with her own string and announces the instances for which her parity calculation matches Bob's one.
The users will discard both bits in the pair if the announced parities are different.
If the parities are the same, the users keep the first bit of the pairs and form a new shorter string from which they will extract the final key.
To this end, they run classical post-processing procedures such as error correction and privacy amplification.
The amount of privacy amplification needed to securely distil a key depends on the security analysis and the resulting rate equation.
In the following, we list the rate equation adopted for each situation analysed in the experiment.

\smallskip

\noindent\textsc{CAL protocol}. This protocol is analysed in the asymptotic scenario for which $P_Z\approx 1$.
The corresponding SKR equation is the one given for `protocol 3' in \cite{CAL19} and the procedure we use to calculate it is similar to the one described in \cite{MPR+19}.
See also \cite{ZHC+19}.
The SKR is the sum of two separate contributions, calculated from each detector $D_0$ and $D_1$ independently: $R_{\textrm{CAL}}=R_{\textrm{CAL}}^{D_0}+R_{\textrm{CAL}}^{D_1}$.
We write the contribution from $D_0$ as
\begin{equation}\label{SKR_Curty}
  R_{\textrm{CAL}}^{D_0} = Q^z [1 - f_{EC} h(E^z)-h(\overline{e}_{1}^{\textrm{ph}})].
\end{equation}
The SKR pertaining to $D_1$ has a similar expression.
In Eq.~\eqref{SKR_Curty}, $h$ is the binary entropy function, $f_{EC}$ is the error correction factor and $Q^z$ and $E^z$ are the gain and the bit error rate, respectively, of the protocol, measured in the experiment from the $D_0$ clicks when the users announce the $Z$ basis.
The quantity $\overline{e}_{1}^{\textrm{ph}}$ is the upper bound to the phase error rate, for which we have \cite{CAL19}
\begin{align}\label{e1Cur}
\overline{e}_{1}^{\textrm{ph}} = \frac{1}{Q^z}\sum_{j=0,1}\[\sum_{m,n=0}^{N_{\textrm{cut}}} c_{m}^{(j)} c_{n}^{(j)}\sqrt{g_{mn}(\overline{Y}^x_{mn},Y_{\textrm{cut}})}\]^2.
\end{align}
In Eq.~\eqref{e1Cur}, the coefficient $c_{k}^{(0)}$ ($c_{k}^{(1)}$) is defined as $c_{k}^{(0)}=e^{-\mu/2} \mu^{k/2}/\sqrt{k!}$ when the integer $k$ is even (odd) and $0$ otherwise;
$g_{mn}(\overline{Y}^x_{mn},Y_{\textrm{cut}})$ is a function equal to $\overline{Y}^x_{mn}$ if $m+n<Y_{\textrm{cut}}$ and equal to $1$ otherwise;
$Y_{\textrm{cut}},N_{\textrm{cut}}$ are two integers such that $Y_{\textrm{cut}}<N_{\textrm{cut}}$.
In our experiment we set $Y_{\textrm{cut}}=8$ and $N_{\textrm{cut}}=12$.
The quantities $\overline{Y}^x_{mn}$ are upper bounds for the yields obtained when Alice (Bob) sends $m$ ($n$) photons.
These are estimated using a constrained optimisation linear program \cite{MPR+19} similar to the standard decoy state technique~\cite{LMC05,Wang05}, with the difference that the yields have to be maximised rather than minimised to provide the worst-case phase error rate.
In our implementation, we measured all the intensity combinations $uu$, $uv$, $uw$, $vv$, $vw$ and $ww$ to improve the decoy-state estimation.
In parallel to this numerical estimation, we also implemented the analytical estimation given in \cite{ZHC+19} to verify the correctness of our results.

\smallskip

\noindent \textsc{SNS protocol}.
The SKR for this protocol in the asymptotic scenario ($P_Z\approx 1$) can be written as \cite{WYH18,XYJ+20}
\begin{equation}\label{SKR_SNSTFQKD}
 R_{\textrm{SNS}}=\underline{Q}_0 + \underline{Q}_1 [1-h(\overline{e}_{1}^{\textrm{ph}})] - f_{EC} Q^z h(E^z).
\end{equation}
In Eq.~\eqref{SKR_SNSTFQKD}, $Q^z$ and $E^z$ are the gain and the bit error rate, respectively, of the protocol, measured in the experiment.
The 0-photon gain and 1-photon gain in the $Z$ basis are
$\underline{Q}_0 = 2 \epsilon (1-\epsilon) e^{-s} e^{-n} \underline{y}_0$ and
$\underline{Q}_1 = 2 \epsilon (1 - \epsilon) (s e^{-s}e^{-n}+n e^{-n} e^{-s}) \underline{y}_1$,
respectively.
The parameters $\underline{y}_1$ ($\underline{y}_0$) and $\overline{e}_1^{\textrm{ph}}$ are, respectively, the lower bound for the single-photon (zero-photon) yield and the upper bound for the single-photon phase error rate.
These quantities are drawn from the $X$ basis of the protocol using equations similar to the ones seen in decoy-states QKD \cite{Wang05,LMC05,WYH18}.

\smallskip

\noindent \textsc{TWCC protocol}.
With the addition of two-way classical communication (TWCC), the users can improve the quality of their data before performing the standard error correction and privacy amplification operations.
The SKR in the asymptotic scenario for this protocol is \cite{XYJ+20}
\begin{equation}\label{SKR_TWCC}
 R_{\textrm{TWCC}}= \frac{1}{N_0} \{ \tilde{n}_1 [1-h(\tilde{e}_{1}^{\textrm{ph}})] - \textrm{leak}_{\textrm{EC}} \},
\end{equation}
with
%
$\tilde{n}_{1} =  n_1^2/(2 n_t)$ ,
$\tilde{e}_{1}^{\textrm{ph}} = 2 \overline{e}_{1}^{\textrm{ph}} (1-\overline{e}_{1}^{\textrm{ph}})$,
$\textrm{leak}_{\textrm{EC}} =  f_{EC} [ n_a h(E_a) + n_b h(E_a) + n_c h(E_c)]$.
%
Here, $n_1=N_0 \underline{Q}_1$ is the number of untagged bits, i.e. the number of bits generated by Charlie's detections when the users send out single-photon states in the $Z$ basis.
$n_t = N_0 Q^z$ is the number of successful detections, an observable of the protocol, with $N_0$ the total number of prepared states.
The term `$\textrm{leak}_{\textrm{EC}}$' represents the number of bits to be exchanged during the error correction procedure.
The quantities $n_a$ and $E_a$ are the number of bits and the error rate, respectively, in Bob's string associated with an odd parity when paired during the TWCC procedure.
Similarly, the quantities $n_b$ and $E_b$ ($n_c$ and $E_c$) are the number of bits and the error rate, respectively, in Bob's string associated with an even parity and when both bits are 0 (1), when paired during the TWCC procedure.
The other quantities are as in Eq.~\eqref{SKR_SNSTFQKD}.

\smallskip

\noindent \textsc{Finite-size SNS and TWCC}.
The finite size analysis of TWCC \cite{CZL+20} is derived directly from the one of SNS~\cite{YHJ+19,JYHW19}.
The error correction term of the asymptotic rate equation \eqref{SKR_TWCC} remains unchanged but the remaining terms are modified to take into account the leakage of information due to finite-size statistical effects.
The number of secret bits in the finite-size regime after TWCC has been performed is given by
\begin{equation}\label{SKR_TWCC_fs}
 n_{\textrm{TWCC-FS}} = \hat{n}_1 [1-h(\hat{e}_{1}^{\textrm{ph}})] - \textrm{leak}_{\textrm{EC}} - \Delta,
\end{equation}
with $\Delta = \log_2 (2/\epsilon_{\textrm{EC}}) - 2 \log_2 (\sqrt{2} \epsilon_{\textrm{PA}} \hat{\epsilon})$ the finite-size correction term and with $\epsilon_{\textrm{EC}}$, $\epsilon_{\textrm{PA}}$ and $\hat{\epsilon}$ the failure probabilities for error correction, privacy amplification and the choice of the smoothing parameter, respectively.
With the right choice of parameters, our implementation features a security parameter of $2.2\times 10^{-9}$, which is the same as in  \cite{CZL+20}.
The hatted quantities $\hat{n}_1$ and $\hat{e}_{1}^{\textrm{ph}}$ correspond to the tilded quantities in Eq.~\eqref{SKR_TWCC}, but calculated in the finite-size regime using a composable definition of security and the Chernoff bound.
Their detailed expressions can be found in the reference paper \cite{JYHW19}.

\medskip

\textbf{Binary maps generation.}
From experiments of the SNS TF-QKD protocol described, real keys were extracted.
To achieve this, single time-tagged events, acquired in 500~ps windows, were processed individually.
Sifting Charlie's announcements, clicks in the $Z$ basis from both detectors were isolated and concatenated.
They were then used by Alice and Bob to separately generate their own initial key string.
For every photon click recorded in $Z$ basis, Alice (Bob) registers a bit 1~(0) if she (he) had sent a weak-coherent pulse within the time slot and a bit 0~(1) if she (he) had chosen not to send anything.
As a result, they obtain matching bits in the cases where only one user has prepared and sent a pulse and opposite bits if both sent.
The latter, accompanied by dark counts, contributes to the QBER in the key generation basis.
A sample of these initial keys for Alice and Bob are shown in the first two squares of Fig.~\ref{fig:2WCCraw}, in the form of binary maps comprised of 128x128 pixels, for the finite-size measurement taken at 522~km.
Zeroes and ones are represented by white and black pixels respectively.
The white-bias of Alice and black bias of Bob are expected and attributed to the send-send clicks that have the highest occurrence probability and in which Alice will always obtain a 1 while Bob will obtain a 0.

Initial keys were post-processed according to the two-way classical communication method to reduce their initial QBER of 16\% and allow successful QKD at such long distances.
During this process, Bob's bits are randomly paired up and their parity calculated.
The pair positions and resulting parity must be publicly announced so that the procedure can be repeated by Alice who will also announce her results.
The initial keys are then further sifted to include only the first bit of pairs whose parity matched in both users.
For instance, given the SNS encoding in the key generation basis, pairs encoded as `$sn$' by Alice (see Protocols in Methods) in a randomly selected pair will provide a matching parity if paired with bits encoded as `$ns$' by Bob whereas will provide unmatched parity if paired with bits encoded as `$ss$' by Bob.
Although TWCC reduces the length of the secret key, it also significantly reduces the QBER so that the overall signal to noise ratio is increased.
The effect of the process on the 522~km data is shown in the first two rectangles at the bottom of Fig.~\ref{fig:2WCCraw}.
The binary map is reduced in dimension by 70\% to represent the equivalent reduction in the entire bit strings.
The white, black bias is also visibly reduced.
To better depict the QBER reduction the binary maps are bitwise XORed before and after TWCC in the rightmost boxes of Fig.~\ref{fig:2WCCraw}.
Matching and opposite bits are represented by white and black pixels respectively.
In this case, the QBER is reduced by over a factor 4.5, from 16\% to 3.5\%, thus allowing us to extract a secret key at distances up to 605.2~km.

\clearpage
\onecolumngrid
\section*{Supplementary Material for:\\600 km repeater-like quantum communications with dual-band stabilisation}

\begin{center}

{
Mirko Pittaluga$^{1,2 \dagger\ast}$,
Mariella Minder$^{1,3 \dagger}$,
Marco Lucamarini$^{1,4 \star}$,
Mirko Sanzaro$^{1}$,
Robert I. Woodward$^{1}$,
Ming-Jun Li$^{5}$,
Zhiliang Yuan$^{1}$
\& Andrew J. Shields$^{1}$
}

\bigskip

\textit{\small
$^{1}$Toshiba Europe Limited, 208 Cambridge Science Park, Cambridge CB4 0GZ, UK \\
$^{2}$School of Electronic and Electrical Engineering, University of Leeds, Leeds LS2 9JT, UK \\
$^{3}$Department of Engineering, Cambridge University, 9JJ Thomson Avenue, Cambridge CB3 0FA, UK \\
$^{4}$Department of Physics and York Centre for Quantum Technologies, University of York, York YO10 5DD, UK \\
$^{5}$Corning Incorporated, Corning, New York, 14831, USA \\
$^{\dagger}$These authors contributed equally to this work \\
$^{\ast}$mirko.pittaluga@crl.toshiba.co.uk\\
$^{\star}$marco.lucamarini@crl.toshiba.co.uk\\
}
\end{center}

\bigskip

\twocolumngrid

\section{Asymmetrical experimental setup}
\begin{figure}[h!]
    \centering
    \includegraphics[width=0.475\textwidth]{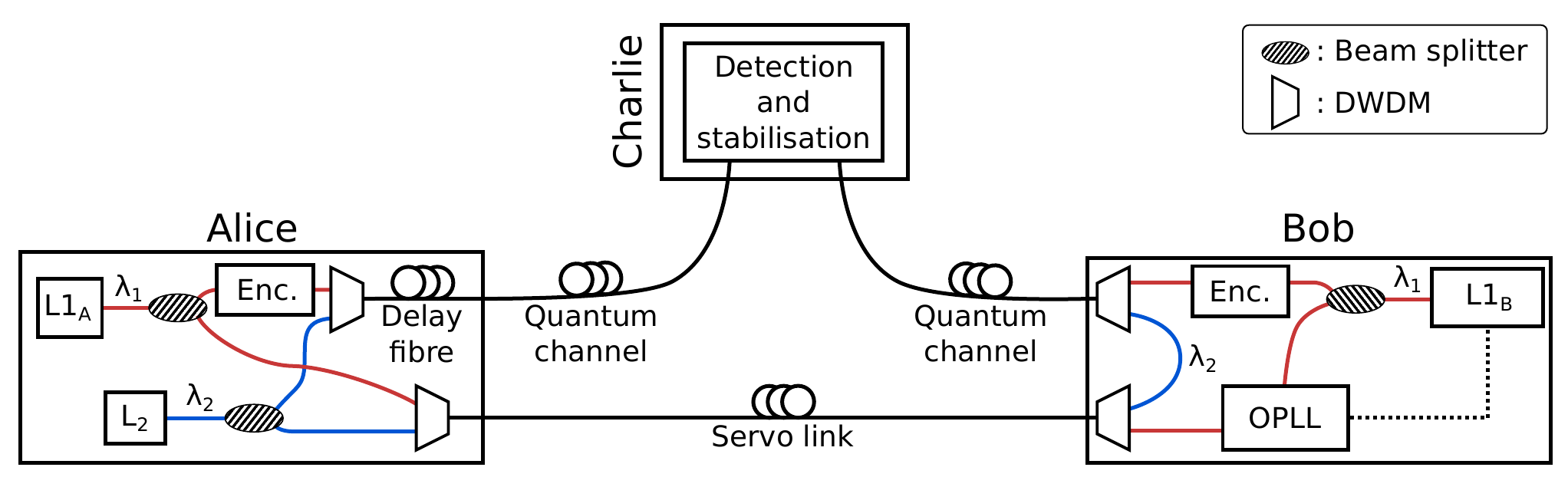}
    \caption{\textbf{Asymmetric TF QKD setup.}
     Alice disseminates $\lambda_1$ and $\lambda_2$ wavelengths to Bob. Alice has a delay fibre equalising the servo fibre length.
     \ZY{DWDM: dense wavelength division multiplexer or demultiplexer, OPLL: optical phase locked loop,} Enc.: encoder, L1\textsubscript{A} (L1\textsubscript{B}): Laser of Alice (Bob),  L\textsubscript{2}: laser for bright reference.
    }
    \label{fig:exp_layout}
\end{figure}

In Fig. 1 in the main text, the central node (Charlie) owns two lasers and disseminates their optical frequencies to the communicating users (Alice and Bob) in a symmetrical manner. Asymmetrical dissemination is also possible, as shown in Fig.~\ref{fig:exp_layout}.  Here, Alice owns two lasers and disseminates their optical frequencies to Bob. The servo fibre becomes a direct link between the users and can bypass Charlie completely.  Asymmetry in the optical channels to Charlie can be compensated by inserting a delay fibre to Alice.

An asymmetrical setup benefits from a simpler central node and requiring just one OPLL. We used this asymmetrical configuration consisting of a 51~km servo and 51~km delay fibre for rapid feasibility studies, before adopting the symmetrical setup with 611~km servo fibre in the final experiments.

\section{Block diagram representation of the feedback systems}
\begin{figure}[!ht]
  \centering
  \includegraphics[width=0.475\textwidth]{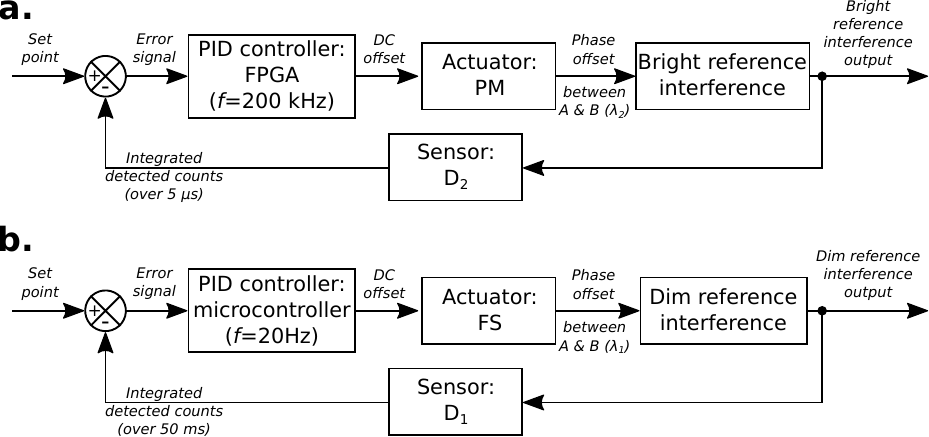}
 \caption{
    \textbf{Block diagrams of the phase-stabilisation system. }
    \emph{Top:}~Feedback executed on $\lambda_2$.
    \emph{Bottom:}~Feedback executed on $\lambda_1$.
    }
  \label{fig:feedback_blocks}
\end{figure}

In Fig.~\ref{fig:feedback_blocks} is reported a block diagram representation of the feedback systems used for the phase stabilisation.
These systems are discussed in detail in the \textbf{Setup} and \textbf{Methods} sections of the main text.
Figure~\ref{fig:feedback_blocks}a shows the stabilisation method based on the bright reference at $\lambda_2$.
Figure~\ref{fig:feedback_blocks}b shows the stabilisation mechanism that corrects the residual phase drift on $\lambda_1$.

\section{System characterisation}
Tab.~\ref{table:sys_param} \ZY{reports} a detailed characterisation of the experimental setup.
This table highlights an asymmetry between the users' insertion losses at Charlie.
The asymmetry is related to the two different types of modulators acting on the incoming signals.
A phase modulator is lossier than a fibre stretcher (which is almost lossless), and therefore Charlie's system transmission for Bob is lower than for Alice.
In the protocol simulations, we used the lowest transmission figure to characterise the losses at the receiver.

In Tab.~\ref{table:sys_param} we present two different dark counts figures.
The first is derived from the SNSPDs (Single Quantum EOS 410 CS cooled at 2.9~K) characterisation, executed with no connected fibre.
The second is extracted from the experimental data during protocol execution.
We associate the increase in dark counts in the second case mostly to the finite DWDM optical isolation between $\lambda_1$ and $\lambda_2$ channels, and to the inelastic scattering occurring in the fibres.
This figure could be further reduced by using spectral filters with better isolation.

\begin{table}[!ht]
    \centering
    \setlength{\tabcolsep}{3mm}
    \begin{tabular}{l|l}
       Charlie's system transmission (from Alice)   & 62.86\% \\
       Charlie's system transmission (from Bob)     & 50.77\% \\
       Efficiency SNSPD D$_0$                       & 73\% \\
       Efficiency SNSPD D$_1$                       & 77\% \\
       SNSPD dark counts (calibration)              & 10 Hz \\
       SNSPD dark counts (experiment)               & 14 Hz \\
       Clock rate for quantum signal                & 500 MHz \\
    \end{tabular}
    \caption{
        Experimental parameters of the setup.
        }
    \label{table:sys_param}
\end{table}

In Tab.~\ref{table:fibre_losses} we report the combined losses of the fibre spools used \ZY{to achieve different quantum channel distances}.
At all distances, we have assigned the lossier spools to Bob.
During the experiment, when there was an asymmetry between the users' photon rate received by Charlie, we compensated for it by increasing the attenuation of Bob's transmitter.

\begin{table}[!ht]
    \centering
    \setlength{\tabcolsep}{3mm}
    \begin{tabular}{l|ll}
         & \multicolumn{2}{c}{Losses (dB)} \\
        Fibre length & Alice  & Bob \\ \hline
        76.641 km &  13.30 & 13.25 \\
        184.351 km & 32.20 & 31.39 \\
        260.866 km & 45.39 & 44.70 \\
        277.461 km & 48.46 & 47.73 \\
        302.585 km & 53.13 & 52.38 \\
    \end{tabular}
    \caption{
        Combined losses of the fibre spools used \ZY{in the quantum channels}.
        }
    \label{table:fibre_losses}
\end{table}

\section{Interference optimisation}
In order to minimise the QBER, it is necessary to optimise the interference between the optical fields prepared by Alice and Bob.
Some aspects of this optimisation, such as spectral overlapping (for both $\lambda_1$ and $\lambda_2$) and time synchronisation (for $\lambda_1$), have already been discussed.
In this section we will describe the details of the polarisation optimisation routine of the setup, and the optimisation of the $\lambda_2$ launch power by the users.

\begin{figure}[b!]
  \centering
  \includegraphics[width=0.38\textwidth]{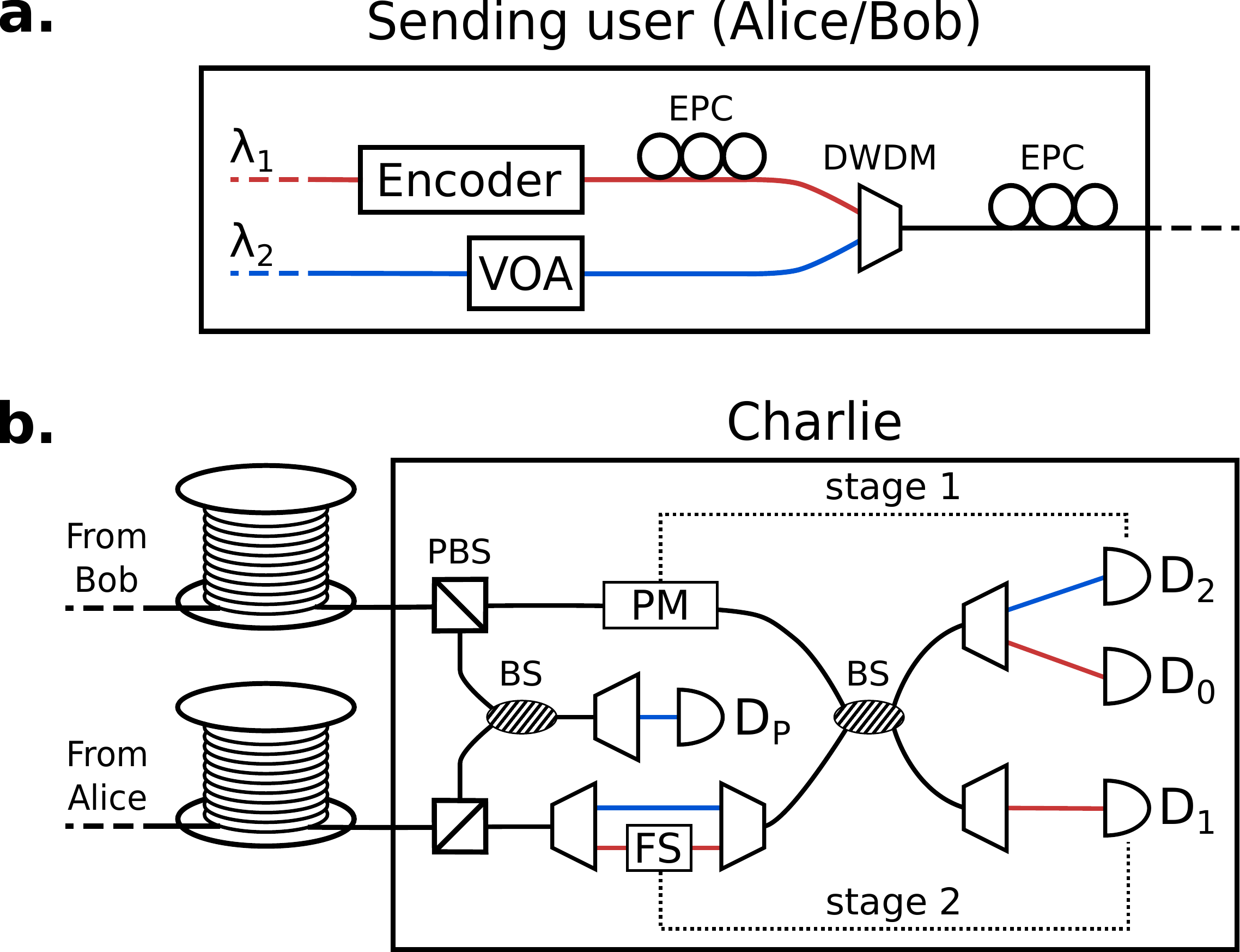}
 \caption{
    \textbf{Details of polarisation and intensity optimisation.}
    \textbf{a.} equipment for $\lambda_1$ and $\lambda_2$ polarisation alignment at the users' stations, and for channel polarisation rotation pre-compensation.
    \textbf{b.} equipment for detecting photons in the wrong polarisation at Charlie.
    \ZY{BS: beam splitter;  D$_P$, D$_0$, D$_1$, D$_2$ : single photon detector; DWDM: dense wavelength division multiplexer/demultiplexer; EPC: electrical polarisation controller; FS: fibre stretcher;} PBS: polarisation beam splitter; VOA: variable optical attenuator.
    }
 \label{fig:interf_optimisation}
\end{figure}
Each user, after executing the encoding over the $\lambda_1$ wavelength, aligns their $\lambda_1$ and $\lambda_2$ signals to the same optical axis through \ZY{an EPC contained in the Encoder boxes (as discussed in the Methods)}.
They then multiplex these signals together, and provide a pre-compensation for the polarisation rotation occurring over the quantum channel through \ZY{a second EPC (see Fig.~\ref{fig:interf_optimisation}a)}.
For this purpose, Charlie is equipped with two PBSs at the entrance of his station, and of an SNSPD, $D_P$ (see Fig.~\ref{fig:interf_optimisation}b), whose purpose is detecting the photons arriving with the wrong polarisation.
Charlie broadcasts the counts received at $D_P$ over a public channel, and the users take turns in changing their EPCs settings with the aim of minimising the detections at $D_P$.
The users' polarisation optimisation routines are run continuously throughout the protocol execution.

In order to optimise the interference of their $\lambda_2$ at Charlie, before running the TF-QKD protocol, each user tunes their $\lambda_2$ launch power with the VOA shown in Fig.~\ref{fig:interf_optimisation}a.

\section{Noise sources in the dual-band stabilisation technique}
The main factors limiting the maximum achievable distance in previous TF-QKD experiments \cite{CZL+20,FZL+20} are the difficulty to stabilise phase over increasingly long communication channels, and the noise associated with the time-multiplexed stabilisation strategy.
The dual-band stabilisation technique provides a solution to both these problems.
It does that by increasing considerably the possible intensity contrast between the stabilisation and the protocol pulses (the contrast is not restricted anymore by the limited range of the intensity modulators, nor by the finite dynamic range of the detectors at Charlie) and by removing double Rayleigh backscattering, the most detrimental source of noise for long-distance time-multiplexed implementations of TF-QKD.

The Rayleigh scattering is a linear process whereby the photons are scattered by the propagation medium.
The scattered photons have the same frequency as the original ones, and introduce noise that cannot be filtered out effectively.
They are in fact uniformly distributed in time (time filtering is not effective), and have the same wavelength as the quantum pulses (spectral filtering is not effective).
In the experiment presented in \cite{CZL+20}, the authors show that in order to maintain a sufficiently intense stabilisation signal, at $\sim$500~km of channel length, the double Rayleigh backscattering starts generating more noise counts than their detectors dark counts.
This noise increases with the launch power of the stabilisation signal, which in turn increases with the channel length.

With the dual-band stabilisation technique, the protocol encoded pulses are at a different wavelength than the stabilisation signal, so that the noise due to the double Rayleigh backscattering can be removed (together with the bright stabilisation signal) through spectral filtering.
The relevant sources of noise with this approach are the channel cross-talk (from the spectral filters) and Raman scattering.

In our experiment the biggest source of noise is the channel cross-talk.
The optical isolation between $\lambda_1$ and $\lambda_2$ is $\sim$52~dB, which introduces $\sim$10~Hz of noise in the protocol analysis when we set 8~MHz target count rate at $D_2$ for the phase stabilisation.
This source of noise could be further reduced using spectral filters with better isolation characteristics, or using two filters in series.
Different from double Rayleigh backscattering, channel cross-talk depends on the intensity of the received signal at Charlie (which has to be increased at most with the square root of the channel length), rather than on the reference signal's launch power (which has to be increased roughly exponentially with the channel length), making this noise source much less dependent on the channel length.

The other scattering-related source of noise in our experiment is Raman scattering, which at 600~km contributes by less than 1 Hz to the noise.
A figure that could be further reduced by using narrower spectral filters.
Similarly to double Rayleigh backscattering, Raman noise increases with the launch power in the channel, but its impact is more than 10 times lower.

\section{Encoding parameters}
In this experiment we tested different TF-QKD protocols: a phase matching version \cite{CAL19}, and the SNS version \cite{WYH18}.
The latter, with and without TWCC (\cite{JYHW19,XYJ+20}).
For all the measurements we used a 25040 pulses long pattern.
Half of the pulses were used for the protocol encoding (dim quantum).
Interleaved to these, phase unmodulated pulses were used as quantum references (dim reference).
The pattern encoding was executed by two time-synchronised pattern generators, which controlled the phase and intensity modulators described in the Methods.
For every protocol, the pattern properties have been chosen with the aim of maximising the communication distance.
With the exception of the point at \ZY{505.2}~km, the two versions of the TWCC SNS protocol (asymptotic and finite size) used the same pattern over the different tested distances.

The way we encode the patterns is by fair sampling.
Using the pattern properties shown in Tab.~\ref{table:pattern_param_for_CAL} and \ref{table:pattern_param_for_SNS}, for each protocol we calculate the probabilities of the different users' pulses combinations.
At this point, we generate a 12520-elements-long list of pulses pairs reflecting the matching probabilities.
This list is then randomly shuffled, resulting in two patterns with a random pulse distribution that respects the matching probabilities expected by the protocol simulations.

Tab.~\ref{table:pattern_param_for_CAL} summarises the photon fluxes and the pattern probabilities used for the CAL TF-QKD protocol.
In this protocol the $X$ basis is the one used for coding, while the $Z$ is used for testing (decoy states).
Pulses in the $X$ basis can only take two phase values: either $+\pi / 2$ or $-\pi / 2$.
Their intensity is set to be equal to the one of the $v$ decoy pulses.
Pulses in the $Z$ basis on the contrary are all phase randomised.
Since this protocol has been run only in the asymptotic regime, for the purpose of the key rate estimation, calculations have been carried out by normalising $P_X$ from the original 50\% to 99.9\%.

\begin{table}
    \centering
    \setlength{\tabcolsep}{3mm}
        $\begin{array}{c|c}
                                & \text{CAL asympt.}    \\
        \text{Fibre length (km)}    & 368.7             \\
        \hline
        s\ \textrm{(ph/pulse)}    & 0.015  \\
        u\ \textrm{(ph/pulse)}    & 0.1    \\
        v\ \textrm{(ph/pulse)}    & 0.015  \\
        w\ \textrm{(ph/pulse)}    & 0.0002 \\
        \hline
        P_Z\ \textrm{(code\ basis)}    & 50.0\% \\
        P_X\ \textrm{(test\ basis)}    & 50.0\% \\
        P_u                   & 33.3\% \\
        P_v                   & 33.3\% \\
        P_w                   & 33.3\% \\
        \hline
        \text{Secret key rate}    & 852.7~\text{bps} \\
        \end{array}$
    \caption{
            Secret key rate and parameters used for the implementation of the CAL TF-QKD protocol~\cite{CAL19} in the asymptotic scenario.
            $s$ is the photon flux used for signal pulses, while $u$, $v$, $w$ are the photon fluxes used for the decoy states.
            $P_Z$ is the users' probability of sending a pulse in the code basis.
            $P_X$ is the users' probability of sending a pulse in the test basis, $P_u$, $P_v$, $P_w$ are the probabilities of sending the $u$, $v$, $w$ decoy pulses respectively.
            }
    \label{table:pattern_param_for_CAL}
\end{table}

Tab.~\ref{table:pattern_param_for_SNS} summarises the parameters for the different configurations of the SNS TF-QKD protocol we tested.
In the SNS protocol, the $Z$ basis is the one used for the key generation (sending-or-not-sending), while the $X$ basis is used for testing the quantum channel (decoy states).
In this protocol, all the encoded pulses (dim quantum) are phase randomised.
Also for SNS, in the asymptotic case the key rate estimation was executed by normalising the probability of using the coding basis $P_Z$ form 50\% to 99.9\%.

\begin{table}
\resizebox{
    0.45\textwidth}{!}{%
    \centering
        $\begin{array}{c|cccc}
                                & \text{SNS asympt.}    & \multicolumn{2}{c}{\text{TWCC SNS asympt.}}  & \text{SNS finite size}    \\
        \hline
         \text{Fibre length (km)}    & 368.7             & \text{all except 605} & \text{605}      & \text{all}                \\
        \hline
         s\ \textrm{(ph/pulse)}    & 0.35   & 0.35   & 0.38    & 0.4    \\
         u\ \textrm{(ph/pulse)}    & 0.35   & 0.35   & 0.38    & 0.4    \\
         v\ \textrm{(ph/pulse)}    & 0.035  & 0.0105 & 0.01065 & 0.075  \\
         w\ \textrm{(ph/pulse)}    & 0.0002 & 0.0002 & 0.00023 & 0.0002 \\
        \hline
         P_Z\ \textrm{(code\ basis)}  & 50.0\% & \multicolumn{2}{c}{50.0\%} & 60.0\% \\
         P_s                 & 5.8\%  & \multicolumn{2}{c}{13.0\%} & 7.5\%  \\
         P_X\ \textrm{(test\ basis)}  & 50.0\% & \multicolumn{2}{c}{50.0\%} & 40.0\% \\
         P_u                 & 33.3\% & \multicolumn{2}{c}{33.3\%} & 20.0\% \\
         P_v                 & 33.3\% & \multicolumn{2}{c}{33.3\%} & 60.0\% \\
         P_w                 & 33.3\% & \multicolumn{2}{c}{33.3\%} & 20.0\% \\
        \end{array}$
    }
    \caption{
            Parameters of the patterns used for the implementation of the SNS TF-QKD protocol \cite{WYH18} in different experimental scenarios (with or without TWCC \cite{JYHW19,XYJ+20}, in the asymptotic or finite-size regimes).
            $s$ is the photon flux used for signal pulses, while $u$, $v$, $w$ are the photon fluxes used for the decoy states.
            $P_Z$ is the users' probability of choosing to encode a pulse in the code basis, $P_s$ is the probability of actually sending a signal when the $Z$ basis is chosen.
            $P_X$ is the users' probability of encoding a pulse in the test basis, $P_u$, $P_v$, $P_w$ are the probabilities of sending the $u$, $v$, $w$ decoy pulses respectively.
            }
    \label{table:pattern_param_for_SNS}
\end{table}

\newpage

\onecolumngrid
\section{Experimental results for SNS-type protocols}
In Tab.
\ref{table:exp_results_SNS_asympt},
\ref{table:exp_results_TWCC_asympt},
\ref{table:exp_results_TWCC_fs},
\ZY{
\ref{table:exp_results_TWCC_asympt_servo-611km} and
\ref{table:exp_results_TWCC_finite-size_611km-servo}},
we report the detailed experimental results for the different SNS-type protocols tested.
\ZY{Data in Tab.~\ref{table:exp_results_SNS_asympt}, \ref{table:exp_results_TWCC_asympt} and \ref{table:exp_results_TWCC_fs} are obtained with the asymmetrical setup (Fig.~\ref{fig:exp_layout}) featuring one OPLL and 51~km servo fibre link.
Data in Tab.~\ref{table:exp_results_TWCC_asympt_servo-611km} and \ref{table:exp_results_TWCC_finite-size_611km-servo} are collected with the symmetrical setup (Fig. 1 in the main text) consisting of 611~km servo fibre link.
}
Tab.~\ref{table:exp_results_SNS_asympt} is reporting the results for the asymptotic SNS protocol (without TWCC).
Tab.~\ref{table:exp_results_TWCC_asympt} \ZY{and \ref{table:exp_results_TWCC_asympt_servo-611km}} summarises the results for the TWCC SNS TF-QKD protocol in the asymptotic scenario, while the ones in the finite size regime are reported in Tab.~\ref{table:exp_results_TWCC_asympt} \ZY{and \ref{table:exp_results_TWCC_finite-size_611km-servo}}.

In all the tables are reported the distances at which the experiments were executed and the number of quantum pulses that were sent ($N_0$).
Also reported are the errors in the different bases and for different pulses intensities, and the calculated secret key rate (S\ZY{KR}) obtained at that distance (alongside the relative SKC$_0$ bound).
For all the protocols, the exact number of pulses sent in each pulses pair configuration can be calculated by multiplying the respective configuration probabilities (deducible from Tab.~\ref{table:pattern_param_for_CAL} and \ref{table:pattern_param_for_SNS}) by $N_0$.
The number of pulses detected in these configurations are reported in the tables below, labelled in the format $B_1B_2t_1t_2$, where $B_1$ and $t_1$ ($B_2$ and $t_2$) are the basis and the type of pulse sent by Alice (Bob).
When TWCC is employed, quantities relative to the key post-processing are listed.

\bigskip

\begin{table}[!hb]
    $
    \begin{array}{c|c}
     \text{\ZY{Quantum link length (km)}} & 368.702 \\
    \hline
     \text{\ZY{Servo link length (km)}} & 51.220 \\
     \hline
     N_0 & 2.066\cdot 10^{11} \\
     \text{Phase mismatch acceptance} & \text{22.5${}^{\circ}$} \\
    \hline
     \text{Z error rate} & \text{6.59$\%$} \\
     \text{Xuu error rate} & \text{3.29$\%$} \\
     \text{Xvv error rate} & \text{3.87$\%$} \\
     \text{Phase error rate} & \text{4.15$\%$} \\
    \hline
     \text{SKR SNS (no TWCC) asympt norm (bit/signal)} & 1.098\cdot 10^{-6} \\
     \text{SKR SNS (no TWCC) asympt norm (bit/s)} & 5.492\cdot 10^2  \\
     \text{Ratio SKR over SKC$_0$} &  1.54  \\
    \hline
     \text{SKC$_0$ (bit/signal)} & 7.151\cdot 10^{-7} \\
     \text{SKC$_0$ (bit/s)} & 3.576\cdot 10^2 \\
    \hline
     \text{Total Detected $D_0$} & 4624363 \\
     \text{Total Detected $D_1$} & 4887891 \\
    \hline
     \text{Detected ZZss} & 39403 \\
     \text{Detected ZZsn} & 314309 \\
     \text{Detected ZZns} & 304872 \\
     \text{Detected ZZnn} & 4264 \\
     \text{Detected ZXsu} & 217790 \\
     \text{Detected ZXsv} & 121824 \\
     \text{Detected ZXsw} & 112334 \\
     \text{Detected ZXnu} & 1729304 \\
     \text{Detected ZXnv} & 173107 \\
     \text{Detected ZXnw} & 1634 \\
     \text{Detected XZus} & 217780 \\
     \text{Detected XZun} & 1786996 \\
     \text{Detected XZvs} & 117638 \\
     \text{Detected XZvn} & 155240 \\
     \text{Detected XZws} & 113964 \\
     \text{Detected XZwn} & 1486 \\
     \text{Detected XXuu} & 1240351 \\
     \text{Detected XXuv} & 685682 \\
     \text{Detected XXuw} & 643480 \\
     \text{Detected XXvu} & 668424 \\
     \text{Detected XXvv} & 115296 \\
     \text{Detected XXvw} & 55695 \\
     \text{Detected XXwu} & 628786 \\
     \text{Detected XXwv} & 62043 \\
     \text{Detected XXww} & 552 \\
    \hline
     \text{Detected XXuu matching ($D_0$)} & 74844 \\
     \text{Detected XXuu matching ($D_1$)} & 79037 \\
     \text{Correct XXuu matching ($D_0$)} & 72352 \\
     \text{Correct XXuu matching ($D_1$)} & 76474 \\
    \hline
     \text{Detected ZZ errors} & 43667 \\
     \text{Detected ZZ correct} & 619181 \\
     \hline
    \end{array}
    $
    \caption{Asymptotic SNS: experimental results  for a \ZY{quantum link} length of 368.7~km \ZY{ and a servo link length of 51~km.}}
    \label{table:exp_results_SNS_asympt}
\end{table}

\begin{table}
        $
        \begin{array}{c|ccccc}
             \text{\ZY{Quantum link length (km)}} & 153.282 & 368.702 & 521.982 & 555.172 & 605.170 \\
            \hline
            \text{\ZY{Servo link length (km)}} & 51.220  & 51.220 & 51.220 & 51.220 & 51.220\\
            \hline
             N_0 & 5.296\cdot 10^{10} & 1.527\cdot 10^{11} & 5.208\cdot 10^{11} & 2.554\cdot 10^{11} & 1.002\cdot 10^{12} \\
             \text{Phase mismatch acceptance} & \text{22.5${}^{\circ}$} & \text{22.5${}^{\circ}$} & \text{22.5${}^{\circ}$} & \text{22.5${}^{\circ}$} & \text{22.5${}^{\circ}$} \\
            \hline
             \text{Z error rate (before)} & \text{13.1$\%$} & \text{13.1$\%$} & \text{14.$\%$} & \text{14.6$\%$} & \text{16.4$\%$} \\
             \text{Odd pairs in raw keys} & 3.522\cdot 10^6 & 2.277\cdot 10^5 & 3.968\cdot 10^4 & 9.565\cdot 10^3 & 1.369\cdot 10^4 \\
             \text{Even pairs 00 in raw keys} & 1.938\cdot 10^6 & 1.228\cdot 10^5 & 2.162\cdot 10^4 & 5.174\cdot 10^3 & 7.411\cdot 10^3 \\
             \text{Even pairs 11 in raw keys} & 1.74\cdot 10^6 & 1.15\cdot 10^5 & 1.963\cdot 10^4 & 4.725\cdot 10^3 & 6.575\cdot 10^3 \\
             \text{Error pairs in raw keys} & 1.591\cdot 10^5 & 1.039\cdot 10^4 & 2.096\cdot 10^3 & 5.532\cdot 10^2 & 1.029\cdot 10^3 \\
             \text{Z error rate (after)} & \text{2.21$\%$} & \text{2.23$\%$} & \text{2.59$\%$} & \text{2.84$\%$} & \text{3.72$\%$} \\
            \hline
             \text{Xuu error rate} & \text{2.8$\%$} & \text{3.21$\%$} & \text{3.86$\%$} & \text{3.68$\%$} & \text{3.5$\%$} \\
             \text{Xvv error rate} & \text{5.53$\%$} & \text{6.32$\%$} & \text{4.78$\%$} & \text{8.33$\%$} & \text{13.6$\%$} \\
             \text{Phase error rate} & \text{5.68$\%$} & \text{6.4$\%$} & \text{3.71$\%$} & \text{6.08$\%$} & \text{2.31$\%$} \\
            \hline
             n_1\text{ (before TWCC)} & 1.159\cdot 10^7 & 7.49\cdot 10^5 & 1.254\cdot 10^5 & 3.284\cdot 10^4 & 3.733\cdot 10^4 \\
             n_1\text{ (after TWCC)} & 3.605\cdot 10^6 & 2.326\cdot 10^5 & 3.685\cdot 10^4 & 1.04\cdot 10^4 & 9.136\cdot 10^3 \\
             e_1{}^{\text{ph}}\text{ (before TWCC)} & \text{5.68$\%$} & \text{6.4$\%$} & \text{3.71$\%$} & \text{6.08$\%$} & \text{2.31$\%$} \\
             e_1{}^{\text{ph}}\text{ (after TWCC - asympt)} & \text{10.7$\%$} & \text{12.$\%$} & \text{7.15$\%$} & \text{11.4$\%$} & \text{4.52$\%$} \\
            \hline
             \text{SKR TWCC asympt norm (bit/signal)} & 7.441\cdot 10^{-5} & 1.412\cdot 10^{-6} & 8.557\cdot 10^{-8} & 2.838\cdot 10^{-8} & 1.937\cdot 10^{-9} \\
             \text{SKR TWCC asympt norm (bit/s)} & 3.721\cdot 10^4 & 7.059\cdot 10^2 & 4.278\cdot 10^1 & 1.419\cdot 10^1 & 9.685\cdot 10^{-1} \\
             \text{Ratio SKR over SKC$_0$} & 0.0215 & 1.97 & 50. & 61.3 & 29.7 \\
            \hline
             \text{SKC$_0$ (bit/signal)} & 3.456\cdot 10^{-3} & 7.151\cdot 10^{-7} & 1.711\cdot 10^{-9} & 4.632\cdot 10^{-10} & 6.511\cdot 10^{-11} \\
             \text{SKC$_0$ (bit/s)} & 1.728\cdot 10^6 & 3.576\cdot 10^2 & 8.556\cdot 10^{-1} & 2.316\cdot 10^{-1} & 3.256\cdot 10^{-2} \\
            \hline
             \text{Total Detected $D_0$} & 66051719 & 4241991 & 745698 & 181085 & 263766 \\
             \text{Total Detected $D_1$} & 69221704 & 4524664 & 795437 & 193156 & 278746 \\
            \hline
             \text{Detected ZZss} & 2423210 & 156953 & 27876 & 6727 & 9620 \\
             \text{Detected ZZsn} & 8146041 & 521082 & 91891 & 22168 & 32219 \\
             \text{Detected ZZns} & 8052039 & 526583 & 91497 & 22122 & 31538 \\
             \text{Detected ZZnn} & 11389 & 1380 & 2024 & 848 & 2907 \\
             \text{Detected ZXsu} & 6196623 & 403188 & 70879 & 17076 & 24785 \\
             \text{Detected ZXsv} & 3217021 & 205870 & 36367 & 8629 & 12876 \\
             \text{Detected ZXsw} & 3107274 & 198690 & 35299 & 8610 & 12320 \\
             \text{Detected ZXnu} & 20581340 & 1349062 & 235655 & 57121 & 81690 \\
             \text{Detected ZXnv} & 604012 & 37884 & 7635 & 1915 & 3125 \\
             \text{Detected ZXnw} & 4212 & 518 & 747 & 289 & 1085 \\
             \text{Detected XZus} & 6202770 & 403207 & 70119 & 17168 & 24722 \\
             \text{Detected XZun} & 20928197 & 1340008 & 236339 & 57217 & 82636 \\
             \text{Detected XZvs} & 3181382 & 208453 & 36111 & 8811 & 12607 \\
             \text{Detected XZvn} & 628269 & 41662 & 7104 & 1879 & 3037 \\
             \text{Detected XZws} & 3078177 & 202140 & 35286 & 8426 & 12396 \\
             \text{Detected XZwn} & 4250 & 534 & 765 & 281 & 1102 \\
             \text{Detected XXuu} & 15759404 & 1022280 & 179333 & 43000 & 61766 \\
             \text{Detected XXuv} & 8202418 & 526279 & 92761 & 22151 & 32573 \\
             \text{Detected XXuw} & 7977865 & 511560 & 89706 & 21900 & 31585 \\
             \text{Detected XXvu} & 8127297 & 531576 & 92456 & 22577 & 31672 \\
             \text{Detected XXvv} & 480868 & 30579 & 5527 & 1427 & 1904 \\
             \text{Detected XXvw} & 247965 & 16325 & 2809 & 799 & 1182 \\
             \text{Detected XXwu} & 7874039 & 516027 & 89699 & 22269 & 31572 \\
             \text{Detected XXwv} & 235716 & 14614 & 2968 & 718 & 1170 \\
             \text{Detected XXww} & 1645 & 201 & 282 & 113 & 423 \\
            \hline
             \text{Detected XXuu matching ($D_0$)} & 930327 & 59658 & 10558 & 2581 & 3712 \\
             \text{Detected XXuu matching ($D_1$)} & 990284 & 64565 & 11389 & 2746 & 3939 \\
             \text{Correct XXuu matching ($D_0$)} & 905683 & 57844 & 10179 & 2478 & 3594 \\
             \text{Correct XXuu matching ($D_1$)} & 961135 & 62388 & 10921 & 2653 & 3789 \\
            \hline
             \text{Detected ZZ errors} & 2434599 & 158333 & 29900 & 7575 & 12527 \\
             \text{Detected ZZ correct} & 16198080 & 1047665 & 183388 & 44290 & 63757 \\
            \hline
        \end{array}
    $
    \caption{Asymptotic TWCC SNS: experimental results obtained \ZY{with a 51~km servo link for various quantum link} fibre lengths.}
    \label{table:exp_results_TWCC_asympt}
\end{table}

\begin{table}
        $
        \begin{array}{c|cccc}
             \text{\ZY{Quantum link length (km)}} & 153.282 & 368.702 & 521.982 & 555.172 \\
             \hline
            \text{\ZY{Servo link length (km)}} & 51.220  & 51.220 & 51.220 & 51.220 \\
            \hline
             N_0 & 6.000\cdot 10^{11} & 2.435\cdot 10^{12} & 3.070\cdot 10^{12} & 3.536\cdot 10^{12} \\
             \text{Phase mismatch acceptance} & \text{22.5${}^{\circ}$} & \text{22.5${}^{\circ}$} & \text{22.5${}^{\circ}$} & \text{22.5${}^{\circ}$} \\
            \hline
             \text{Z error rate (before)} & \text{7.67$\%$} & \text{7.69$\%$} & \text{9.01$\%$} & \text{9.77$\%$} \\
             \text{Odd pairs in raw keys} & 4.164\cdot 10^7 & 3.564\cdot 10^6 & 2.333\cdot 10^5 & 1.336\cdot 10^5 \\
             \text{Even pairs 00 in raw keys} & 2.135\cdot 10^7 & 1.838\cdot 10^6 & 1.198\cdot 10^5 & 6.922\cdot 10^4 \\
             \text{Even pairs 11 in raw keys} & 2.077\cdot 10^7 & 1.768\cdot 10^6 & 1.152\cdot 10^5 & 6.501\cdot 10^4 \\
             \text{Error pairs in raw keys} & 5.738\cdot 10^5 & 4.938\cdot 10^4 & 4.542\cdot 10^3 & 3.103\cdot 10^3 \\
             \text{Z error rate (after)} & \text{0.685$\%$} & \text{0.689$\%$} & \text{0.97$\%$} & \text{1.16$\%$} \\
            \hline
             \text{Xuu error rate} & \text{2.69$\%$} & \text{2.88$\%$} & \text{2.87$\%$} & \text{3.47$\%$} \\
             \text{Xvv error rate} & \text{3.56$\%$} & \text{3.81$\%$} & \text{5.31$\%$} & \text{5.09$\%$} \\
             \text{Phase error rate} & \text{4.16$\%$} & \text{4.47$\%$} & \text{6.04$\%$} & \text{5.59$\%$} \\
            \hline
             n_1\text{ (before TWCC)} & 1.185\cdot 10^8 & 1.003\cdot 10^7 & 6.4\cdot 10^5 & 3.652\cdot 10^5 \\
             n_1\text{ (after TWCC)} & 3.596\cdot 10^7 & 3.013\cdot 10^6 & 1.828\cdot 10^5 & 1.026\cdot 10^5 \\
             e_1{}^{\text{ph}}\text{ (before TWCC)} & \text{4.23$\%$} & \text{4.71$\%$} & \text{7.48$\%$} & \text{7.65$\%$} \\
             e_1{}^{\text{ph}}\text{ (after TWCC)} & \text{8.09$\%$} & \text{8.98$\%$} & \text{13.8$\%$} & \text{14.1$\%$} \\
            \hline
             \text{Number of secure bits generated (bits)} & 1.707\cdot 10^7 & 1.329\cdot 10^6 & 4.046\cdot 10^4 & 1.745\cdot 10^4 \\
             \text{SKR (bit/signal)} & 2.846\cdot 10^{-5} & 5.459\cdot 10^{-7} & 1.318\cdot 10^{-8} & 4.937\cdot 10^{-9} \\
             \text{SKR (bit/s)} & 1.423\cdot 10^4 & 2.729\cdot 10^2 & 6.59 & 2.468 \\
             \text{Ratio SKR over SKC$_0$} & 0.00823 & 0.763 & 7.7 & 10.7 \\
            \hline
             \text{SKC$_0$ (bit/signal)} & 3.456\cdot 10^{-3} & 7.151\cdot 10^{-7} & 1.711\cdot 10^{-9} & 4.632\cdot 10^{-10} \\
             \text{SKC$_0$ (bit/s)} & 1.728\cdot 10^6 & 3.576\cdot 10^2 & 8.556\cdot 10^{-1} & 2.316\cdot 10^{-1} \\
            \hline
             \text{Total Detected $D_0$} & 601407532 & 50532067 & 3317070 & 1945930 \\
             \text{Total Detected $D_1$} & 623261177 & 54266217 & 3616047 & 2071809 \\
            \hline
             \text{Detected ZZss} & 14322977 & 1229675 & 80689 & 46061 \\
             \text{Detected ZZsn} & 90153430 & 7739935 & 511622 & 296484 \\
             \text{Detected ZZns} & 90035335 & 7685653 & 507468 & 290238 \\
             \text{Detected ZZnn} & 642059 & 54944 & 20176 & 17462 \\
             \text{Detected ZXsu} & 25714805 & 2199489 & 144929 & 82891 \\
             \text{Detected ZXsv} & 46330397 & 3956642 & 262219 & 150458 \\
             \text{Detected ZXsw} & 12870272 & 1107377 & 72977 & 42765 \\
             \text{Detected ZXnu} & 159718589 & 13637478 & 898693 & 517464 \\
             \text{Detected ZXnv} & 90956994 & 7710475 & 507227 & 297300 \\
             \text{Detected ZXnw} & 93804 & 7989 & 2973 & 2550 \\
             \text{Detected XZus} & 25704406 & 2209257 & 146032 & 85067 \\
             \text{Detected XZun} & 159575781 & 13695784 & 906734 & 526484 \\
             \text{Detected XZvs} & 46091214 & 3944702 & 259209 & 149050 \\
             \text{Detected XZvn} & 89990485 & 7718247 & 507676 & 294097 \\
             \text{Detected XZws} & 12982067 & 1109205 & 73179 & 42204 \\
             \text{Detected XZwn} & 92445 & 7776 & 2977 & 2516 \\
             \text{Detected XXuu} & 44724005 & 3838050 & 252841 & 145782 \\
             \text{Detected XXuv} & 82278922 & 7047247 & 466076 & 270776 \\
             \text{Detected XXuw} & 22992576 & 1978960 & 131444 & 75985 \\
             \text{Detected XXvu} & 82116355 & 7042214 & 463506 & 268561 \\
             \text{Detected XXvv} & 78177798 & 6684280 & 437657 & 252988 \\
             \text{Detected XXvw} & 13010771 & 1116305 & 73487 & 42281 \\
             \text{Detected XXwu} & 22952199 & 1964545 & 129678 & 75082 \\
             \text{Detected XXwv} & 13127709 & 1110997 & 73238 & 42795 \\
             \text{Detected XXww} & 13314 & 1058 & 410 & 398 \\
            \hline
             \text{Detected XXuu matching ($D_0$)} & 2822546 & 239368 & 15656 & 9148 \\
             \text{Detected XXuu matching ($D_1$)} & 2880087 & 250202 & 16753 & 9609 \\
             \text{Correct XXuu matching ($D_0$)} & 2743311 & 232285 & 15215 & 8846 \\
             \text{Correct XXuu matching ($D_1$)} & 2805790 & 243186 & 16263 & 9260 \\
            \hline
             \text{Detected ZZ errors} & 14965036 & 1284619 & 100865 & 63523 \\
             \text{Detected ZZ correct} & 180188765 & 15425588 & 1019090 & 586722 \\
            \hline
        \end{array}
        $
    \caption{Finite-size TWCC SNS: experimental results obtained \ZY{with a servo link length of 51~km} at various \ZY{quantum link} fibre lengths.}
    \label{table:exp_results_TWCC_fs}
\end{table}


\begin{table}
        $
   \begin{array}{c|cc}
 \text{Quantum link length (km)} & 555.172 & 605.170 \\
\hline
  \text{Servo link length (km)} & 611.448 & 611.448 \\
  \hline
 N_0 & 8.038 \cdot 10^{11} & 1.353 \cdot 10^{12} \\
 \text{Phase mismatch acceptance} & \text{22.5${}^{\circ}$} & \text{22.5${}^{\circ}$} \\
\hline
 \text{Z error rate (before)} & \text{14.1$\%$} & \text{16.3$\%$} \\
 \text{Odd pairs in raw keys} & 2.622 \cdot 10^4 & 1.631 \cdot 10^4 \\
 \text{Even pairs 00 in raw keys} & 1.445 \cdot 10^4 & 8.530 \cdot 10^3 \\
 \text{Even pairs 11 in raw keys} & 1.272 \cdot 10^4 & 8.134 \cdot 10^3 \\
 \text{Error pairs in raw keys} & 1.412 \cdot 10^3 & 1.205 \cdot 10^3 \\
 \text{Z error rate (after)} & \text{2.64$\%$} & \text{3.65$\%$} \\
\hline
 \text{Xuu error rate} & \text{4.99$\%$} & \text{5.41$\%$} \\
 \text{Xvv error rate} & \text{9.68$\%$} & \text{13.$\%$} \\
 \text{Phase error rate} & \text{5.12$\%$} & \text{5.94$\%$} \\
\hline
 n_1\text{ (before TWCC)} & 8.615 \cdot 10^4 & 5.346 \cdot 10^4 \\
 n_1\text{ (after TWCC)} & 2.631 \cdot 10^4 & 1.5756 \cdot 10^4 \\
 e_1{}^{\text{ph}}\text{ (before TWCC)} & \text{5.12$\%$} & \text{5.94$\%$} \\
 e_1{}^{\text{ph}}\text{ (after TWCC - asympt)} & \text{9.71$\%$} & \text{11.2$\%$} \\
\hline
 \text{SKR TWCC asympt norm (bit/signal)} & 2.936 \cdot 10^{-8} & 1.555 \cdot 10^{-9} \\
 \text{SKR TWCC asympt norm (bit/s)} & 1.468 \cdot 10^1 & 7.778 \cdot 10^{-1} \\
 \text{Ratio SKR over }\text{SKC}_0 & 63.4 & 24.0 \\
\hline
 \text{SKC}_0\text{ (bit/signal)} & 4.632 \cdot 10^{-10} & 6.468 \cdot 10^{-11} \\
 \text{SKC}_0\text{ (bit/s)} & 2.316 \cdot 10^{-1} & 3.234 \cdot 10^{-2} \\
\hline
 \text{Total Detected }D_0 & 494716 & 312237 \\
 \text{Total Detected }D_1 & 523903 & 329627 \\
\hline
 \text{Detected ZZss} & 18133 & 11390 \\
 \text{Detected ZZsn} & 61210 & 37648 \\
 \text{Detected ZZns} & 59884 & 38261 \\
 \text{Detected ZZnn} & 1822 & 3392 \\
 \text{Detected ZXsu} & 46849 & 28773 \\
 \text{Detected ZXsv} & 24278 & 14942 \\
 \text{Detected ZXsw} & 23458 & 14489 \\
 \text{Detected ZXnu} & 153672 & 96970 \\
 \text{Detected ZXnv} & 4919 & 4045 \\
 \text{Detected ZXnw} & 685 & 1319 \\
 \text{Detected XZus} & 46598 & 28882 \\
 \text{Detected XZun} & 157244 & 96571 \\
 \text{Detected XZvs} & 23398 & 14948 \\
 \text{Detected XZvn} & 4967 & 4004 \\
 \text{Detected XZws} & 22942 & 14598 \\
 \text{Detected XZwn} & 686 & 1371 \\
 \text{Detected XXuu} & 118175 & 73243 \\
 \text{Detected XXuv} & 62210 & 38161 \\
 \text{Detected XXuw} & 60417 & 36867 \\
 \text{Detected XXvu} & 60505 & 38517 \\
 \text{Detected XXvv} & 3671 & 2582 \\
 \text{Detected XXvw} & 2015 & 1588 \\
 \text{Detected XXwu} & 58686 & 37257 \\
 \text{Detected XXwv} & 1940 & 1548 \\
 \text{Detected XXww} & 255 & 498 \\
 \left.\text{Detected XXuu matching (}D_0\right) & 6963 & 4178 \\
 \left.\text{Detected XXuu matching (}D_1\right) & 7452 & 4358 \\
 \left.\text{Correct XXuu matching (}D_0\right) & 6624 & 3974 \\
 \left.\text{Correct XXuu matching (}D_1\right) & 7071 & 4100 \\
 \text{Detected ZZ errors} & 19955 & 14782 \\
 \text{Detected ZZ correct} & 121094 & 75909 \\
 \hline
\end{array}
               $
    \caption{\ZY{Asymptotic TWCC SNS with 611~km servo link: experimental results obtained at various quantum link fibre lengths.}}
    \label{table:exp_results_TWCC_asympt_servo-611km}
\end{table}

\begin{table}
        $
       \begin{array}{c|c}
 \text{Quantum link length (km)} & 555.172 \\
\hline
 \text{Servo link length (km)} & 611.448\\
 \hline
 N_0 & 3.121 \cdot 10^{12} \\
 \text{Phase mismatch acceptance} & \text{22.5${}^{\circ}$} \\
\hline
 \text{Z error rate (before)} & \text{9.98$\%$} \\
 \text{Odd pairs in raw keys} & 9.163 \cdot 10^4 \\
 \text{Even pairs 00 in raw keys} & 4.608 \cdot 10^4 \\
 \text{Even pairs 11 in raw keys} & 4.594 \cdot 10^4 \\
 \text{Error pairs in raw keys} & 2.234 \cdot 10^3 \\
 \text{Z error rate (after)} & \text{1.22$\%$} \\
\hline
 \text{Xuu error rate} & \text{5.12$\%$} \\
 \text{Xvv error rate} & \text{5.49$\%$} \\
 \text{Phase error rate} & \text{5.73$\%$} \\
\hline
 n_1\text{ (before TWCC)} & 2.597 \cdot 10^5 \\
 n_1\text{ (after TWCC)} & 7.530 \cdot 10^4 \\
 e_1{}^{\text{ph}}\text{ (before TWCC)} & \text{8.32$\%$} \\
 e_1{}^{\text{ph}}\text{ (after TWCC)} & \text{15.2$\%$} \\
\hline
 \text{Number of secure bits generated (bits)} & 1.109 \cdot 10^4 \\
 \text{SKR (bit/signal)} & 3.555 \cdot 10^{-9} \\
 \text{SKR (bit/s)} & 1.777 \\
 \text{Ratio SKR over }\text{SKC}_0 & 7.675 \\
\hline
 \text{SKC}_0\text{ (bit/signal)} & 4.632 \cdot 10^{-10} \\
 \text{SKC}_0\text{ (bit/s)} & 2.316 \cdot 10^{-1} \\
\hline
 \text{Total Detected }D_0 & 1330784 \\
 \text{Total Detected }D_1 & 1429497 \\
\hline
 \text{Detected ZZss} & 31819 \\
 \text{Detected ZZsn} & 200662 \\
 \text{Detected ZZns} & 202447 \\
 \text{Detected ZZnn} & 12911 \\
 \text{Detected ZXsu} & 57243 \\
 \text{Detected ZXsv} & 103341 \\
 \text{Detected ZXsw} & 28818 \\
 \text{Detected ZXnu} & 357386 \\
 \text{Detected ZXnv} & 209380 \\
 \text{Detected ZXnw} & 1823 \\
 \text{Detected XZus} & 57723 \\
 \text{Detected XZun} & 354724 \\
 \text{Detected XZvs} & 102785 \\
 \text{Detected XZvn} & 203294 \\
 \text{Detected XZws} & 29110 \\
 \text{Detected XZwn} & 1825 \\
 \text{Detected XXuu} & 99664 \\
 \text{Detected XXuv} & 183501 \\
 \text{Detected XXuw} & 51213 \\
 \text{Detected XXvu} & 183158 \\
 \text{Detected XXvv} & 175638 \\
 \text{Detected XXvw} & 29287 \\
 \text{Detected XXwu} & 51513 \\
 \text{Detected XXwv} & 30769 \\
 \text{Detected XXww} & 247 \\
 \left.\text{Detected XXuu matching (}D_0\right) & 5842 \\
 \left.\text{Detected XXuu matching (}D_1\right) & 6417 \\
 \left.\text{Correct XXuu matching (}D_0\right) & 5529 \\
 \left.\text{Correct XXuu matching (}D_1\right) & 6102 \\
 \text{Detected ZZ errors} & 44730 \\
 \text{Detected ZZ correct} & 403109 \\
 \hline
\end{array}
 $
    \caption{\ZY{Finite-size TWCC SNS with and 555.2~km quantum link fibre and 611~km servo fibre.}}
    \label{table:exp_results_TWCC_finite-size_611km-servo}
\end{table}


\end{document}